\documentclass[%
reprint,
nofootinbib,
amsmath,amssymb,
aps,
prc,
]{revtex4-1}

\usepackage{graphicx}
\usepackage{dcolumn}
\usepackage{bm}
\usepackage{pifont}
\usepackage{subfiles}
\usepackage{units}
\usepackage[breaklinks=true]{hyperref}
\usepackage{booktabs}
\begin{document}
	
	\newcommand*\rfrac[2]{{}^{#1}\!/_{#2}}
	\setlength{\textfloatsep}{20pt plus 5.0pt minus 3.0pt}
	
	\preprint{APS/123-QED}
	
	\title{Heavy ion anisotropies: a closer look at the angular power spectrum}
	
	\author{M. Machado$^{1,2}$}
	\affiliation{%
		Niels Bohr International Academy$^1$ and Discovery Center$^2$, Niels Bohr Institute, Blegdamsvej 17, DK-2100 Copenhagen, Denmark
	}%

	\date{\today}
	
	\begin{abstract}
		
		Anisotropies in the final state of heavy-ion collisions carry information on the creation, expansion and evolution of the quark-gluon plasma. Currently, there is an abundance of studies on azimuthal anisotropies in comparison to longitudinal ones. The purpose of this work is to quantify angular $(\theta, \phi)$ correlations to further the understanding of the full spatial 3-D picture of emitted hadrons. Therefore, public ALICE data from Run 1 (2010) of Pb-Pb collisions at $\sqrt{s_{NN}} = 2.76\unit{~TeV}$ is analyzed through the estimation of an angular power spectrum. Issues with $|\eta| < 0.9$ limitation are tackled, as well as event multiplicity and detector efficiency. Firstly, spectra are calculated for toy Monte Carlo samples. Secondly, heavy-ion data spectra are presented for the full momentum phase space $0.15 < p_T < 100\unit{~GeV}$ and also separate intervals $p_T < 0.54\unit{~GeV}$ and $p_T > 0.54\unit{~GeV}$. The latter reveal how different geometries dominate at distinct scales and transverse momentum. Finally, the study submits particles generated through the AMPT model to the same power spectrum analysis. This comparison shows that in scales dominated by flow geometry, AMPT qualitatively describes the data spectra, while the opposite is true for smaller scales.     
		
	\end{abstract}
	
	\maketitle
	
	
	\section{\label{sec:level1}Introduction}
	
	When two heavy nuclei collide in the Large Hadron Collider (LHC) at ultrarelativistic energies, they form a state of matter denoted quark-gluon plasma (QGP)~\cite{qgp_form:jet_quenching1,qgp_form:sup_Y,qgp_form:dijet} which, as the name indicates, has quarks and gluons as its degrees of freedom. Such is believed to be the state of the universe itself when it was microseconds old. After being formed, the QGP undergoes a collective expansion, i.e., the initial geometry molded by the overlap region and quantum fluctuations dictates the emitted particles final distribution. As the system cools, the quarks and gluons change into hadronic matter through a smooth phase transition. A hadron gas is thus formed and its components finally reach the detectors: these particles carry the only information available on the properties and behavior of this hot primordial soup.
	
	From cosmology, our older window into the early universe comes from the photons emitted from the surface of last scattering, when electrons were bound to nuclei and formed atoms. As the universe expanded, the wavelengths of said photons also stretched. This electromagnetic radiation was first discovered in 1964 by radio astronomers and it is denoted Cosmic Microwave Background (CMB)~\cite{cmb_first}. The anisotropies present on the latter are an imprint of the initial conditions of our universe. These preferred directions of emission can be studied from the \textit{angular power spectrum} of the CMB, which can determine not only the curvature of the universe, but can be also used to get information on dark-matter and dark-energy densities~\cite{planck_res}.
	
	Studies on the \textit{azimuthal} anisotropies found in the product of heavy-ion collisions have provided a plethora of insights into QGP formation and evolution: fluctuations in the initial geometry, constraints on shear viscosity to entropy ratio $\eta/s$, and the QGP equation of state~\cite{azimuthal1,azimuthal2,azimuthal3}. However, in order to get the full QGP picture, it is necessary to understand its \textit{longitudinal} dynamics. Still, the latter is yet to be addressed with the same level of scrutiny applied to the transverse evolution. For instance, there is evidence of event-by-event fluctuations in pseudorapidity~\cite{longi_ani1,longi_ani2}, whose effects distinct models manage to reproduce only qualitatively~\cite{review1}, thus underlying the importance of 3+1D modeling. Additionally, the dependence of $\eta/s$ on QGP temperature could potentially be probed through longitudinal anisotropies~\cite{Denicol:2015nhu}, as the temperature profile of the medium changes with pseudorapidity.     
	
	An angular power spectrum $C_{\ell}$ of heavy ions aims at enriching the understanding of anisotropies in the final particle distribution as a whole: both in the transverse and longitudinal directions. What is more, pseudorapidity $\eta$ is defined in terms of the polar angle $\theta$ between a particle's 3-momentum and the beam axis through $\eta = -\log(\tan(\theta/2))$. This relation enables for converting particles coordinates $(\eta, \phi) \to (\theta, \phi)$, which allows for power spectrum calculation. Said quantity measures the amplitude of correlations between $(\theta, \phi)$ pairs as a function of angular scale. 
	
	This work is a detailed follow-up on the study in Ref.~\cite{PhysRevC.99.054910}: it also projects heavy-ion ALICE\footnote{A Large Ion Collider Experiment} data~\cite{open_data,cbourjau} onto spherical surfaces and calculates their averaged spectrum $\langle C_{\ell} \rangle$. The first difference lies in the choice of higher map resolution. Additionally, events are separated in accordance to their interaction points, i.e., the position along the beam axis where the collision occurs. The method built in Ref.~\cite{PhysRevC.99.054910} is then applied for each event batch, considering the usual issues of detector limited acceptance and non-uniform efficiency, as well as relatively low multiplicities. Furthermore, the present study extends the power spectrum estimation for a divided transverse momentum $p_T$ phase space: $p_T < 0.54\unit{~GeV}$ and $p_T > 0.54\unit{~GeV}$. Finally, spectra results are compared to a multi-phase transport (AMPT) model~\cite{ampt} for a single centrality class, 10-20\%. A difference in scales is clearly observed, with the model succeeding in one, though failing the other.  
	
	The first part of this paper details the method built in Ref.~\cite{PhysRevC.99.054910} for toy Monte Carlo simulations under the new chosen resolution. The MC distributions are sampled from functions of the type $f_{MC}(\theta, \phi) = g(\theta) h(\phi)$. On the second part, the angular power spectrum of Pb-Pb collisions at $2.76\unit{~TeV}$ center-of-mass energy is presented for the ALICE Run 1 2010 data set from the CERN open data portal~\cite{open_data}. The results are displayed for the full and partial $p_T$ intervals in the following centralities: 0-5\%, 5-10\%, 10-20\%, 20-30\%, and 30-40\%.
	
	
	
	\section{Building a method}
	
	
	Particles emitted from nuclei collisions have their coordinates represented in terms of pseudorapidity $\eta$, azimuthal angle $\phi$ and transverse momentum $p_T$. As previously mentioned, the data set at hand was recorded with the ALICE detector~\cite{Collaboration_2008,alice_2014} in 2010, during Run 1 of LHC at a center-of-mass energy per nucleon of $\sqrt{s_{NN}} = 2.76\unit{~TeV}$~\cite{open_data}. 
	
	The extraction of said ALICE data along with its event selection and default cuts were executed through the repository in Ref.~\cite{cbourjau}. The algorithm begins by verifying whether a primary vertex $z_{vtx}$ exists, i.e., if the particles came from a heavy-ion collision or from vacuum chamber interactions. The next step then consists in checking if $z_{vtx}$ lies within $10\unit{~cm}$ of the detector's center. In addition, event multiplicity should be non-zero. Lastly, a minimum-bias trigger selects high efficiency events~\cite{alice_2014}.
	
	The tasks of primary and secondary vertex determination, as well as track reconstruction, centrality estimation and separation of particle beam from background are performed by the following subsystems: the Inner Tracking System (ITS)~\cite{its}, the Time Projection Chamber (TPC)~\cite{tpc} and the VZERO (V0) detectors~\cite{vzero}. Through the combination of their capabilities and event selection criteria, the phase space coverage at present is $|\eta| < 0.9$, $0.15 < p_T < 100\unit{~GeV}$ and $0 \leq \phi < 2\pi$.
	
	The main objective of this work lies in mapping the heavy-ion events onto tessellated spheres. As a first step, the final particle distribution $f(\eta, \phi, p_T)$ should undergo a change of variables from pseudorapidity to polar angle through the expression $\theta = 2\arctan(e^{-\eta})$; explicitly, $f(\eta, \phi, p_T) \to f(\theta, \phi, p_T)$. The software package HEALPix\footnote{\url{https://healpix.sourceforge.io/}}~\cite{healpix} (Hierarchical Equal Area isoLatitude Pixelation) was employed for both map projections and power spectrum calculations. As the name itself suggests, it divides the surface of a sphere into pixels of equal areas. 
	
	The number of pixels $N_{pix}$ is directly related to the resolution parameter $N_{side}$ through $N_{pix} = 12N^2_{side}$. In the previous study~\cite{PhysRevC.99.054910} the chosen resolution was $N_{side} = 8$. However, due to concerns over signal smoothing of events from most central collisions, the present resolution is $N_{side} = 16$.
	
	During the course of this paper event maps are made by counting the particles with coordinates $(\theta, \phi)$ that fall within each pixel boundaries. More specifically, heavy-ion data or samples of a function $f(\theta, \phi)$ are sets of unit vectors $\mathbf{\hat{n}} = (\theta, \phi)$. Each of their entries $(\theta_j, \phi_j)$ correspond to a pixel on a map: the 2-D angular distribution turns into an 1-D array of pixels indexed $p \in [0, N_{pix})$ from $\theta = 0$ to $\theta = \pi$. Lastly, it should be remarked that $p_T$ will not be considered in this first part of the discussion.
	
	Let $\mathbf{n_p} = (\theta_p, \phi_p)$ be the pixel center coordinates, then $f(\mathbf{n_p})$ represents the pixelated map of $f(\mathbf{\hat{n}})$. For a distribution on the surface of a tessellated sphere, the $a_{\ell m}$ coefficients of the spherical harmonic expansion can be estimated as 
	
	\begin{equation}
	a_{\ell m} = \sum_{p = 0}^{N_{pix} - 1} \Omega_p f(\mathbf{n_p}) Y^*_{\ell m}(\mathbf{n_p}),
	\label{eq:alm_exp}
	\end{equation}
	
	\noindent where $\Omega_p = 4\pi/N_{pix}$ is the standard pixel weight under HEALPix. 
	
	The angular power spectrum $C_{\ell}$ is defined as the variance, or second moment, of the $a_{\ell m}$ for a given $\ell$. The later are denoted \textit{multipole}, with $\ell = 0$ being the monopole, $\ell = 1$ the dipole and so on. The expression for $C_{\ell}$ is
	
	\begin{align}
	\langle a_{\ell m} a^*_{\ell' m'} \rangle = \delta_{\ell \ell'}\delta_{m m'} C_{\ell}&, \nonumber \\
	C_{\ell} = \frac{1}{2\ell + 1}\sum_{m = -\ell}^{\ell} |a_{\ell m}|^2&.
	\label{eq:powspec}
	\end{align}
	
	\noindent Following up, we discuss in detail the effects on the spectrum under limited detector coverage, how to estimate the background caused by low multiplicity and non-uniform detector efficiency for a MC generated distribution.
	
	\subsection{The mask effect}
	
	The acceptance of a detector may be represented by the function $W(\mathbf{\hat{n}})$, also denoted as \textit{mask}. The true underlying distribution $f_{tru}(\mathbf{\hat{n}})$ then relates to the observed one $f_{obs}(\mathbf{\hat{n}})$ through $f_{obs}(\mathbf{\hat{n}}) = W(\mathbf{\hat{n}}) f_{tru}(\mathbf{\hat{n}})$. Also, for a perfect detector $W(\mathbf{\hat{n}}) = 1$, $\forall \mathbf{\hat{n}}$.
	
	Under $W(\mathbf{\hat{n}})$ the $a_{\ell m}$ coefficients of the maps in question, i.e., the ones accessible experimentally, become a linear combination of the coefficients pertaining to the true distributions under full detector coverage $\tilde{a}_{\ell m}$. For the current data set, $W(\mathbf{\hat{n}}) = 1$ if $44^o \lesssim \theta \lesssim 136^o$ and zero otherwise. In addition, the azimuthal direction is completely covered. The harmonic coefficients are related as follows:
	
	\begin{equation}
	a_{\ell m} = \sum_{\ell'}\sum_{m' = -\ell'}^{\ell'} \underbrace{\left[ \int_{\Omega_{\eta}} Y_{\ell' m'}(\mathbf{\hat{n}}) Y^*_{\ell m}(\mathbf{\hat{n}}) d\Omega \right]}_{W^{\ell \ell'}_{m m'}} \tilde{a}_{\ell' m'},
	\label{eq:alm_corr}
	\end{equation} 
	
	\noindent where $\Omega_{\eta}$ is the region covered by the detector.  
	
\begin{figure}[!ht]
		\includegraphics[width=0.48\textwidth]{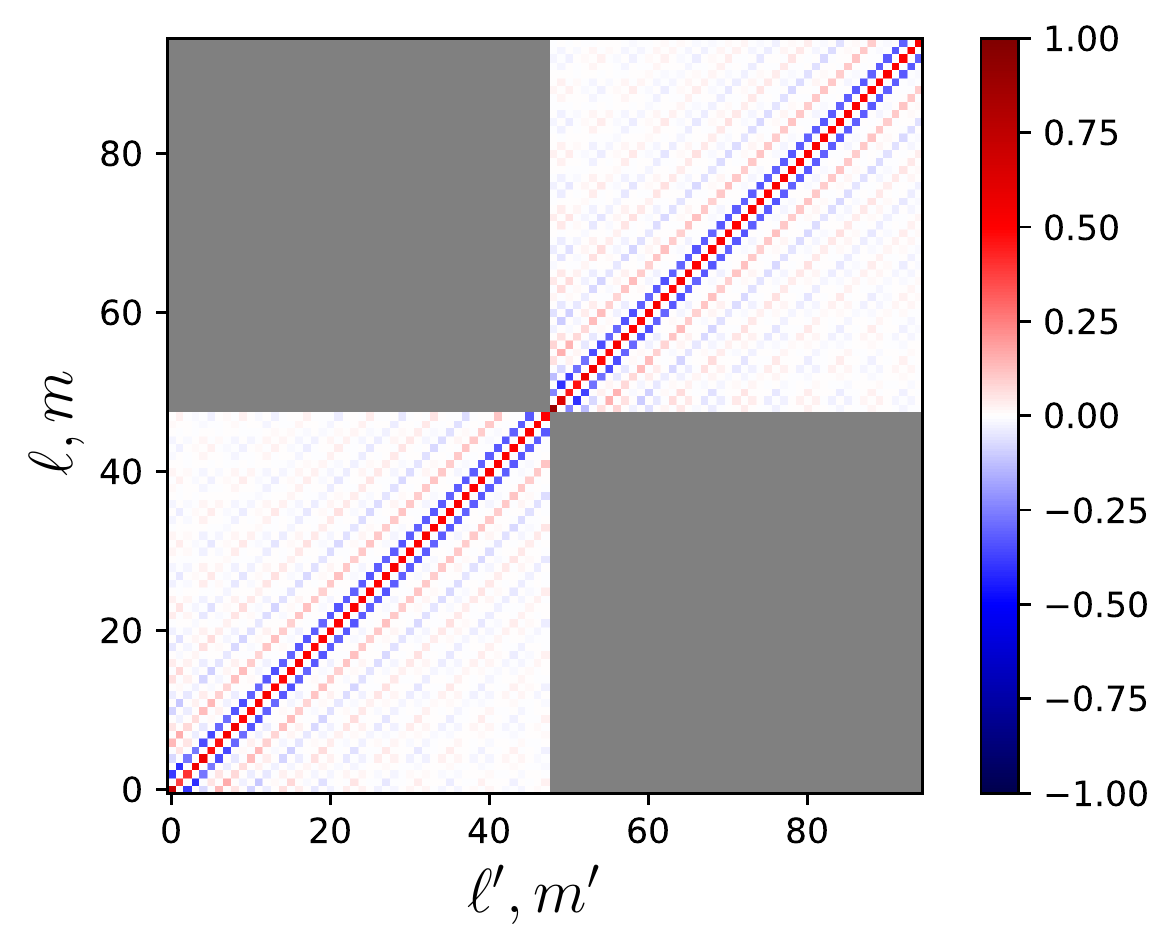}
		\caption{\label{fig:Wllmm} Mixing matrix $W^{\ell \ell'}_{m m'}$ for $m, m' = 0, 1$ and truncated to $\ell_{max} = 47$.}
	\end{figure}

	The mixing matrix $W^{\ell \ell'}_{m m'}$ associated with the mask $W(\mathbf{\hat{n}})$ is depicted in Fig.~\ref{fig:Wllmm} for a maximum multipole value $\ell_{max} = 47$ corresponding to the chosen resolution. Each one of the major squares refers to a fixed combination of $m, m' = 0, 1$ with varying $\ell, \ell'$. The gray areas indicate $m \neq m'$, which yield null values for the matrix, since detector coverage encompasses the full azimuth. Another noteworthy feature of $W^{\ell \ell'}_{m m'}$ is the chessboard-like pattern of $m = m'$: it means that $\ell, \ell'$ must have the same parity. Additionally, the matrix values decrease with increasing $\Delta \ell = |\ell' - \ell|$, so only harmonics with neighboring multipole values contribute to the observed $a_{\ell m}$. 
	
	The linear system in Eq.~(\ref{eq:alm_corr}) has no unique solution, due to $W^{\ell \ell'}_{m m'}$ having a determinant equal to zero. In this case, it is not possible to get the true $\tilde{a}_{\ell m}$ values by inverting the matrix. The solution proposed in Ref.~\cite{PhysRevC.99.054910} relies on the fact that $a_{\ell 0}$ coefficients with $\ell$ even are the ones mostly affected by the mask geometry. The aforementioned work shows how fully isotropic distributions have a particular enhancement of even modes relative to odd ones solely due to $W(\mathbf{\hat{n}})$. It is straightforward to verify with Eq.~(\ref{eq:alm_corr}) that $\tilde{a}_{0 0}$ alongside $W^{\ell \ell'}_{m m'}$ make it so that only $a_{\ell 0}$ with $\ell$ even are non-zero. Accordingly, all remaining modes should be zero, which provides an interesting asset to the method at hand. 
	
	From Ref.~\cite{PhysRevC.99.054910}, anisotropies are best accounted for when $m = 0$ modes are eliminated from $C_{\ell}$. The following expression for the power spectrum is hence used throughout the analysis:
	
	\begin{equation}
	C^{m\neq0}_{\ell} = \frac{1}{2\ell+1} \sum_{m=-\ell}^{\ell} |a_{\ell m}|^2 - \frac{|a_{\ell 0}|^2}{2\ell+1}.
	\label{eq:powspec_mdz}
	\end{equation}
	
	\subsection{The multiplicity issue}
	
	The averaged angular power spectrum $C^{m\neq0}_{\ell}$ of isotropic maps limited to $|\eta| < 0.9$ should be equal to zero for all $\ell$. Nevertheless, for generated isotropic distributions with multiplicity akin to those of the 0-5\% centrality, their averaged spectrum $\langle C^{m\neq0}_{\ell} \rangle_{iso}$ is of order $\mathcal{O}(10^{-3})$~\cite{PhysRevC.99.054910}. This suggests that each $\ell$-mode of the spectra carries a quantity attributed to the distributions' multiplicity $M$.  
	
	In order to quantify how the observed spectrum values change with typical event multiplicity, a simple approach was devised: we generated a set of 8000 isotropic events where they all have the same multiplicity and calculated $\langle C_{\ell}^{m\neq0} \rangle_{iso}$. This process is repeated from $M = 100$ to $M = 5000$. Additionally, the value of $\langle C^{m\neq0}_{\ell} \rangle_{iso}$ for each $\ell$ is plotted as a function of the multiplicity set and fitted to a power law of the type $p_0\cdot M^{-p_1} + p_2$, where $p_i$ for $i=0,1,2$ are the fit parameters. 
	
	\begin{figure}[!ht]
		\includegraphics[width=0.48\textwidth]{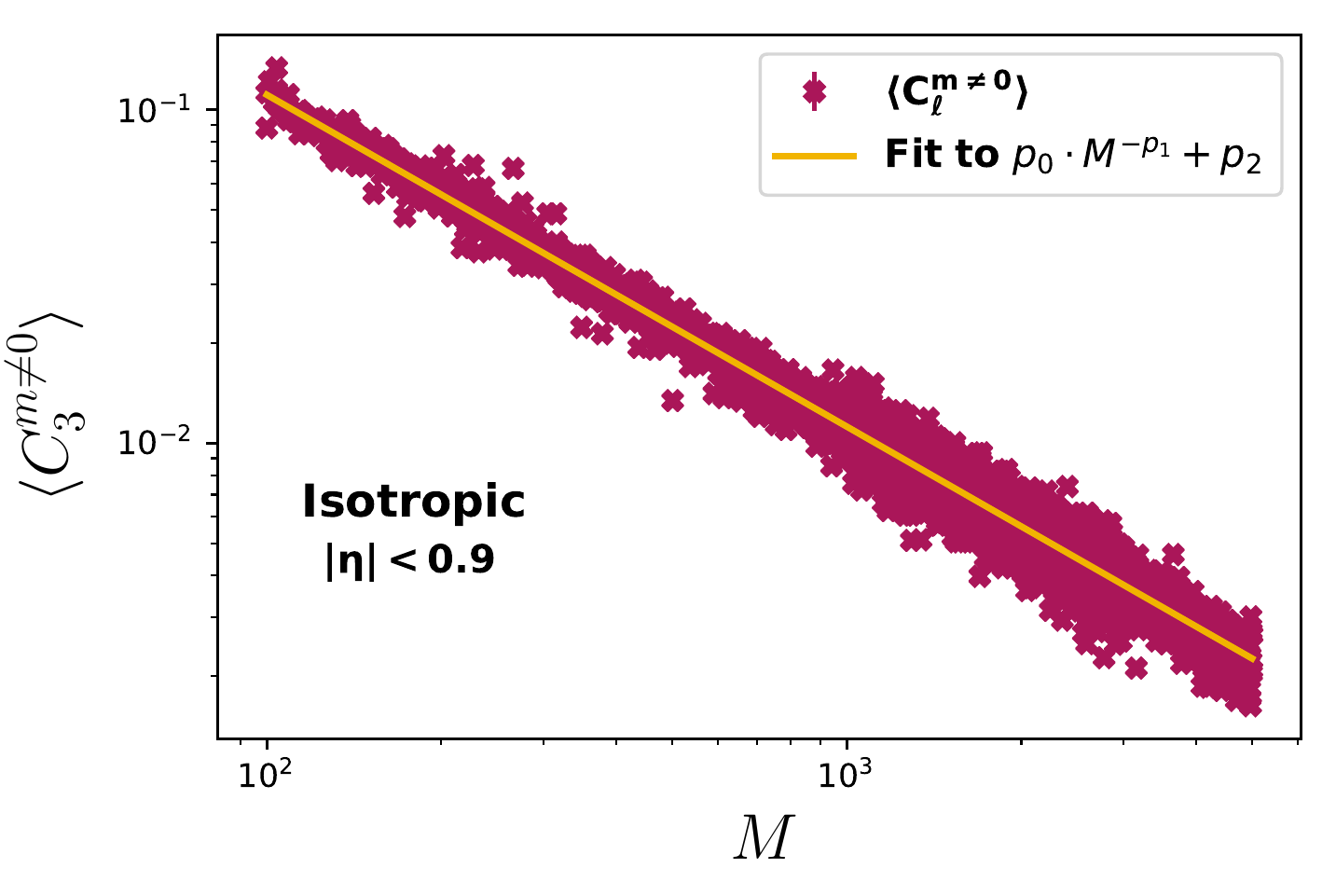}
		\caption{\label{fig:c3vsmult} $\langle C_3^{m\neq0} \rangle_{iso}$ as a function of multiplicity ($M$).}
	\end{figure}
	
	Notice in Fig.~\ref{fig:c3vsmult} how $\langle C_3^{m\neq0} \rangle_{iso}$ decreases with higher multiplicity according to a power law, a feature that is true to all modes. It is worth noticing that $p_2$ should be the `true' $\langle C_{\ell}^{m\neq0} \rangle$ value when multiplicity tends to infinity. One could be tempted in this case to correct the spectrum values by finding $p_0, p_1$ and subtracting $\langle p_0 \cdot M^{-p_1} \rangle$ from the observed $\langle C^{m\neq0}_{\ell} \rangle$. However, the relation in Fig.~\ref{fig:c3vsmult} could be dependent on the underlying distribution. Instead, we make use of the non-zero result of $\langle C_{\ell}^{m\neq0} \rangle_{iso}$ itself for a given multiplicity distribution.
	
	Let $\langle N^{m\neq0}_{\ell} \rangle$ be the averaged background spectrum associated with the typical event multiplicity. We begin by producing $\sim 10^6$ isotropic events according to the present multiplicity distribution within $|\eta| < 0.9$ and calculating their power spectra. At the end, the ensemble average is taken thus yielding $\langle N^{m\neq0}_{\ell} \rangle$. This power spectrum should give an estimate on the sparsity background, i.e. the size of fluctuations originated solely from typical event multiplicity. 
	
	A simple approach to correcting the averaged power spectrum of a given set of events would be to subtract from it the corresponding $\langle N^{m\neq0}_{\ell} \rangle$ to its multiplicity. This can be thought as akin to comparing the signal-to-background ratio to unity $|\langle C^{m\neq0}_{\ell} \rangle / \langle N^{m\neq0}_{\ell} \rangle - 1|$. Since the only difference is a normalization factor, we define $\langle S^{m\neq0}_{\ell} \rangle = |\langle C^{m\neq0}_{\ell} \rangle - \langle N^{m\neq0}_{\ell} \rangle|$ the averaged angular power spectrum corrected by the low event multiplicity.
	
	In order to test the efficacy of this correction, two sets of Monte Carlo distributions are generated. Their underlying functions have the form $f_{MC_1}(\mathbf{\hat{n}}) = h(\phi)$ and $f_{MC_2}(\mathbf{\hat{n}}) = g(\theta)h(\phi)$, with 
	
	\begin{align}
	g(\theta) &\propto \cosh\left(\frac{\theta - \pi/2}{2}\right)((\theta - \pi/2)^2 + 1), \nonumber \\
	h(\phi) &\propto \left[1 + 2\sum_{n=1}^{6}v_n\cos(n(\phi - \Psi_n))\right],
	\label{eq:mc_dist}
	\end{align}
	
	\noindent where $v_n$ are fixed coefficients of the Fourier expansion (shown in Table~\ref{tab:vn_coefs}) and $\Psi_n$ varies randomly event-by-event for each $n$. The same number of events $N_{evts} \sim 8000$ were generated for both mentioned functions, with same multiplicity distributions. These simulated events were then projected onto maps and had their power spectra calculated. Finally, the average was taken resulting in $\langle C^{m\neq0}_{\ell} \rangle_{MC_1}$ and $\langle C^{m\neq0}_{\ell} \rangle_{MC_2}$. The same background spectrum $\langle N^{m\neq0}_{\ell} \rangle$ was used for both their corrections, $\langle S^{m\neq0}_{\ell} \rangle_{MC_1}$ and $\langle S^{m\neq0}_{\ell} \rangle_{MC_2}$.
	
	\begin{table}[!ht]
		\centering
		\begin{tabular}{ *{3}{c} } \hline\hline
			$v_1$ & $v_2$ & $v_3$ \\ \midrule
			\textbf{0.02119} & \textbf{0.05928} & \textbf{0.02636} \\\hline\hline 
			$v_4$ & $v_5$ & $v_6$ \\ \midrule
			\textbf{0.01218} & \textbf{0.00520} & \textbf{0.00209} \\ \hline
		\end{tabular}
		\caption{\label{tab:vn_coefs} Values of $v_n$ coefficients for $h(\phi)$.}	
	\end{table}
	
	\begin{figure}[!ht]
		\includegraphics[width=0.48\textwidth]{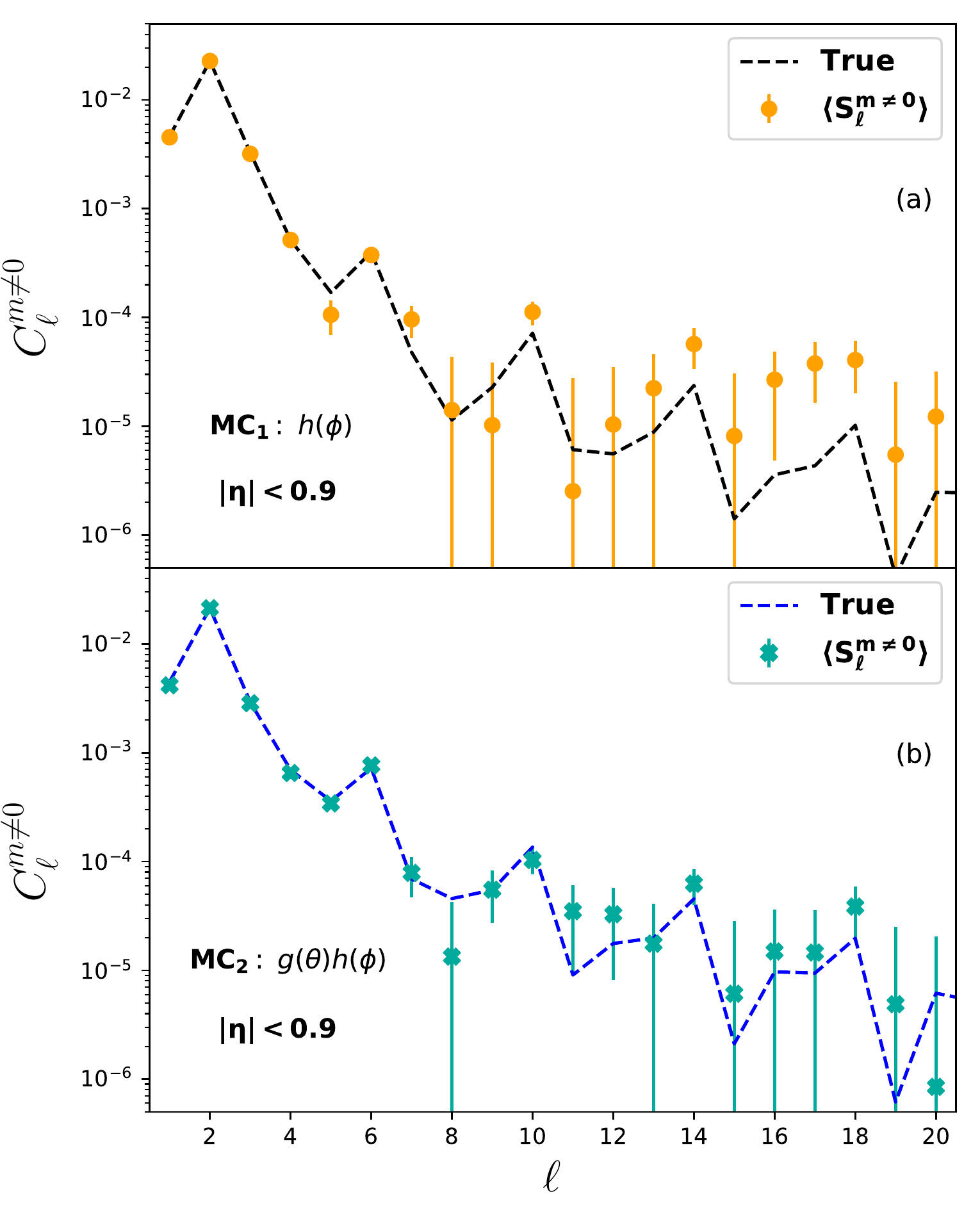}
		\caption{\label{fig:mc_corr_avgcls} Comparison between corrected $\langle S^{m\neq0}_{\ell} \rangle$ and the analytically calculated spectra for both MC functions.}
	\end{figure}
	
	The resulting spectra $\langle S^{m\neq0}_{\ell} \rangle_{MC_1}$ and $\langle S^{m\neq0}_{\ell} \rangle_{MC_2}$ are compared to the analytically calculated ones $\mathtt{C}^{m\neq0}_{\ell}$ directly from $f_{MC_1}(\mathbf{\hat{n}})$ and $f_{MC_2}(\mathbf{\hat{n}})$. These are depicted in Fig.~\ref{fig:mc_corr_avgcls} until $\ell = 20$. Notice how they fit the direct calculation quite closely for $1 \leq \ell \leq 6$. At higher $\ell$ modes the influence of the $v_n$ coefficients decreases with smaller scales. It is an unsurprising outcome, since the geometry generated by initial condition fluctuations belongs to large scales, or low $\ell$.
	
	At this point one can notice from Fig.~\ref{fig:mc_corr_avgcls} how the angular power spectrum of distributions purely dominated by flow should look like. The peak in $\ell = 2$ pertains to the almond-like shape of the overlapping nuclei. It also influences $\ell = 6$, since $v_2$ also contributes to $C_6$. Though their values differ, the shapes of $\langle S^{m\neq0}_{\ell} \rangle_{MC_1}$ and $\langle S^{m\neq0}_{\ell} \rangle_{MC_2}$ are basically the same, specially at the low $\ell$ region.
	
	Overall, the method of correction described above managed to perform remarkably well for $\ell \leq 6$, while becoming trickier for higher values. The limitations of the correction are probably dependent on the distributions at hand, since it is clear to see that $MC_2$ was better estimated than $MC_1$ on the $7 \leq \ell \leq 10$ region. Before moving on to heavy-ion data, it is necessary to ascertain that the strategy applied in Ref.~\cite{PhysRevC.99.054910} to deal with detector efficiency actually works.
	
	\subsection{Detector efficiency}
	
\begin{figure}[!ht]
		\includegraphics[width=0.48\textwidth]{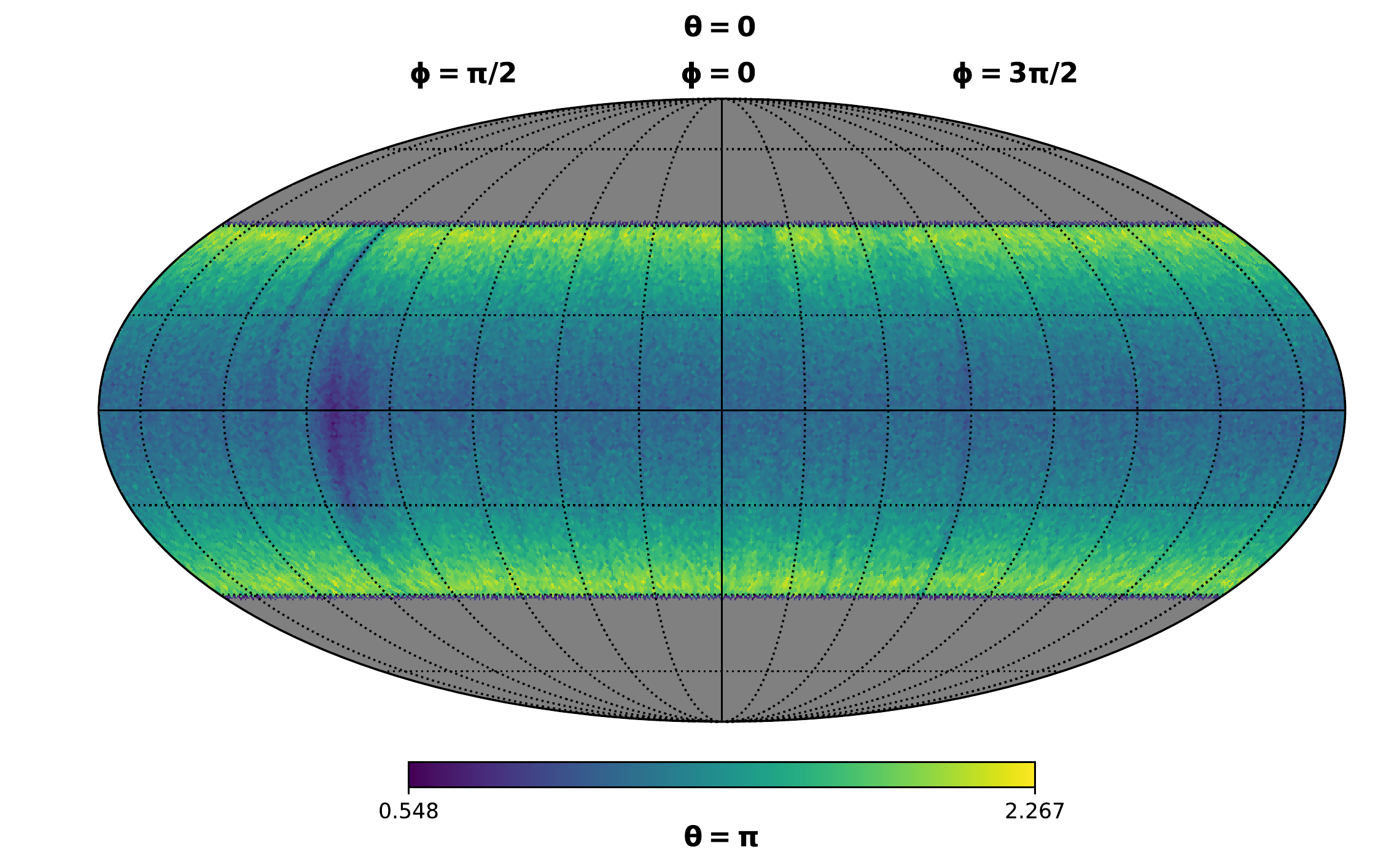}
		\caption{\label{fig:supmap1015} Overlapping map $F^{all}(\mathbf{n_p})$ of $\sim 8000$ events belonging to the 10-15\% centrality.}
	\end{figure}

	The collisions of heavy ions have different azimuthal orientations relative to each other, implying that if they were to be summed over, the result would be isotropic in $\phi$. Namely, $\sum_{i=0}^{N_{evts}-1}f^{(i)}(\mathbf{n_p}) \propto g(\mathbf{n_p})$, with $f^{(i)}(\mathbf{n_p})$ as the map of a single event $i$ and $g(\mathbf{n_p})$ representing the pixelation of $g(\theta)$. Such overlapping of collisions is shown in Fig.~\ref{fig:supmap1015} for the 10-15\% centrality. However, anisotropies are clearly seen, which suggests they are caused by the detector's efficiency. In the following calculations, we consider the $MC_2$ distribution, since it possesses $g(\theta)$, making it closer to the data itself.
	
	Assuming the detector's efficiency to be a function of the polar and azimuthal angles, $D(\mathbf{\hat{n}})$, with corresponding map $D(\mathbf{n_p})$. Since each event is subjected to the same detector performance, then their particle distribution on the sphere $f^{(i)}_{MC_2}(\mathbf{n_p})$ should instead be $f_{D_2}^{(i)}(\mathbf{n_p}) = D(\mathbf{n_p}) \cdot f^{(i)}_{MC_2}(\mathbf{n_p})$, where $f_{D_2}^{(i)}(\mathbf{n_p})$ designates the observed event particle distribution map under detector effects. In light of these new considerations, the overlapping distribution of the observed events will be 
	
	\begin{align}
	F^{all}(\mathbf{n_p}) &= \sum_{i = 0}^{N_{evts}-1} f_{D_2}^{(i)}(\mathbf{n_p}) \nonumber \\ 
	&= D(\mathbf{n_p}) g(\mathbf{n_p}).
	\label{eq:fall}
	\end{align}
	
\begin{figure}[!ht]
		\includegraphics[width=0.48\textwidth]{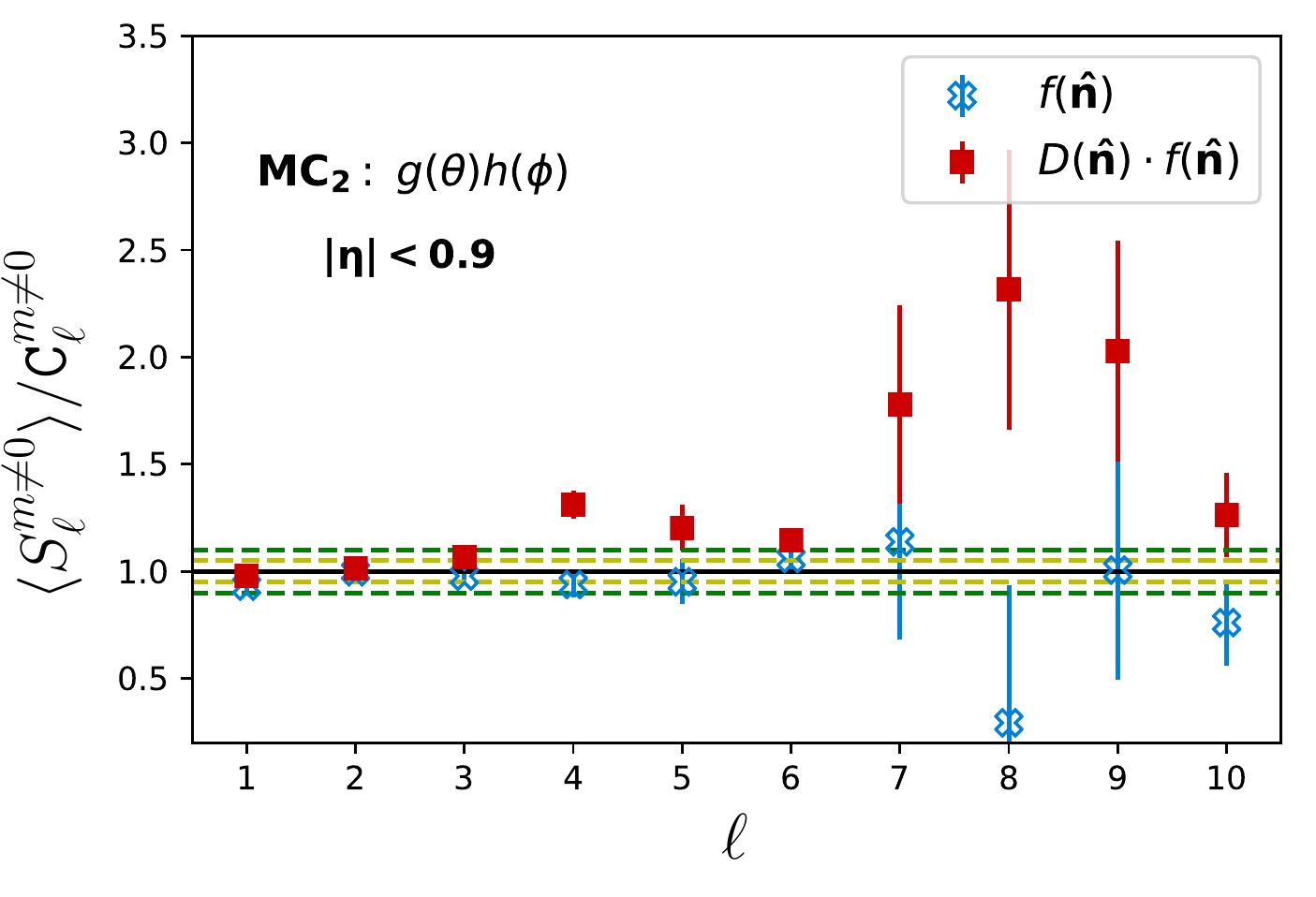}
		\caption{\label{fig:ratios} Ratios $\langle S^{m\neq0}_{\ell} \rangle / \mathtt{C}^{m\neq0}_{\ell}$ for $MC_2$ under detector with uniform and non-uniform efficiencies. Dashed lines are $1\sigma$ (yellow) and $2\sigma$ (green) deviations for $f_{MC_2}(\mathbf{\hat{n}})$ within $1 \leq \ell \leq 6$ from unity.}
	\end{figure}

	In order to verify the changes in spectrum values associated with the detector anisotropies, a 2-D spline of $D(\mathbf{n_p})$ was created to estimate $D(\mathbf{\hat{n}})$. Then, $\sim8000$ events with underlying function $D(\mathbf{\hat{n}})f_{MC_2}(\mathbf{\hat{n}})$ were generated, had their spectra calculated and averaged over. The same background spectrum $\langle N^{m\neq0}_{\ell} \rangle$ applied to $MC_1$ and $MC_2$ above was used to compute $\langle S^{m\neq0}_{\ell} \rangle_{D_2}$. On the follow-up, both corrected spectra $\langle S^{m\neq0}_{\ell} \rangle_{MC_2}$ and $\langle S^{m\neq0}_{\ell} \rangle_{D_2}$ are divided by the analytically calculated $\mathtt{C}^{m\neq0}_{\ell}$ for $MC_2$. These ratios are shown in Fig.~\ref{fig:ratios}. 
	
	Notice in Fig.~\ref{fig:ratios} how, in the region $\ell \leq 6$ where the coefficients $v_n$ have most influence, it is possible to see a deviation at $\ell = 4, 5, 6$. Interestingly, $\ell = 8$ is the most accentuated mode when checking the ratio $\langle S^{m\neq0}_{\ell} \rangle_{D_2} / \mathtt{C}^{m\neq0}_{\ell}$. After all, it corresponds precisely to the scale $\sim 22.5^o$ of the dark patch in Fig.~\ref{fig:supmap1015}.
	
	In the previous work~\cite{PhysRevC.99.054910}, the trick employed to deal with detector-caused anisotropies was dividing each event by $F^{all}(\mathbf{n_p})$. The same approach is used here:
	
	\begin{equation}
	\overline{f}_{D_2}(\mathbf{n_p}) = \frac{f_{D_2}(\mathbf{n_p})}{D(\mathbf{n_p})g(\mathbf{n_p})}.
	\label{eq:fbar}
	\end{equation}
	
	\noindent Note how the latter results are also divided by their multiplicity density distribution in $\theta$. Therefore, besides the changes on $\langle C_l \rangle$ due to smoothing out the detector anisotropies, there should be the additional effect of changing the overall $\theta$ distribution. In other words, the pixels in $\overline{f}_{D_2}(\mathbf{n_p})$ will have associated weights coming from both detector efficiencies and the distribution along the polar angle. Nevertheless, the pixel density will remain unchanged, as the ones with null value shall stay like that.
	
	A direct consequence of dividing by $g(\mathbf{n_p})$ lies in the estimation of $\langle N^{m\neq0}_{\ell} \rangle$. A new calculation was thus devised: we take an event, randomize its azimuthal distribution to get rid of the $\phi$ dependency and then divide it by the average event map, $g(\mathbf{n_p})$. This process is repeated $\sim 10^6$ times, each turn taking a random event. Finally the averaged power spectrum is calculated, here denoted $\langle N^{m\neq0}_{\ell} \rangle_{bar}$. These normalized azimuthally isotropic events possess the same trait as $\overline{f}_{D_2}(\mathbf{n_p})$, namely weighted pixels whose densities follow $g(\mathbf{n_p})$.
	
\begin{figure}[!ht]
		\includegraphics[width=0.48\textwidth]{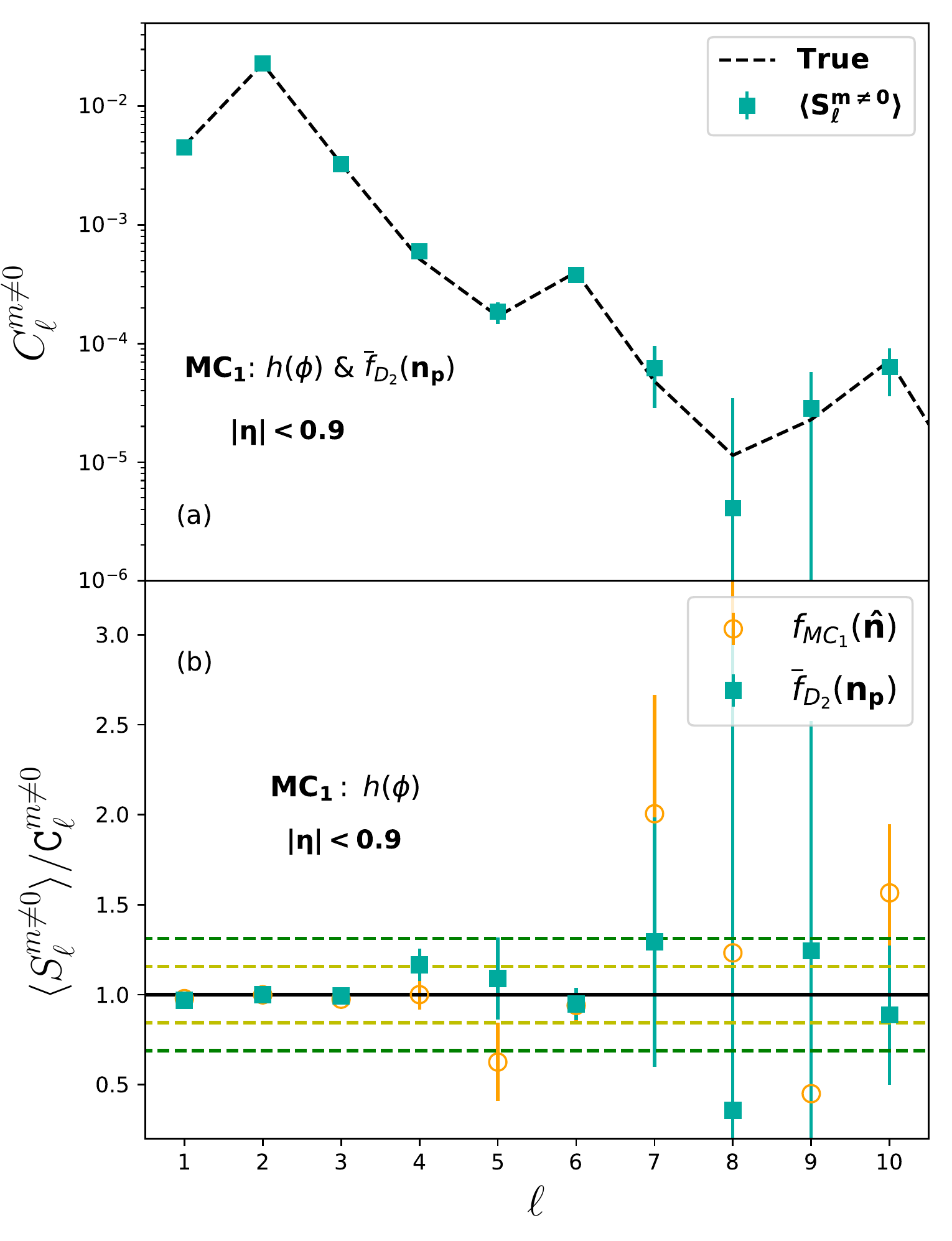}
		\caption{\label{fig:mc2_detmodf} Efficiency-corrected spectrum $\langle S^{m\neq0}_{\ell} \rangle_{bar}$ in comparison to the expected $MC_1$ spectrum (a). Ratios $\langle S^{m\neq0}_{\ell} \rangle/\mathtt{C}^{m\neq0}_{\ell}$ for $f_{MC_1}(\mathbf{\hat{n}})$ and $\overline{f}_{D_2}(\mathbf{n_p})$ (b). Dashed lines are $1\sigma$ (yellow) and $2\sigma$ (green) deviations of the $MC_1$ ratio from unity within $1 \leq \ell \leq 6$.}
	\end{figure}

	It should also be remarked that Eq.~(\ref{eq:fbar}) implies that $\overline{f}_{D_2}(\mathbf{n_p})$ will have a spectrum like $f_{MC_1}(\mathbf{\hat{n}})$, since its polar distribution has been smoothed out. We denote the corrected power spectrum of $\overline{f}_{D_2}(\mathbf{n_p})$ as $\langle S^{m\neq0}_{\ell} \rangle_{bar}$ and compare it to $\mathtt{C}^{m\neq0}_{\ell}$ pertaining to $f_{MC_1}(\mathbf{\hat{n}})$. The result can be seen in Fig.~\ref{fig:mc2_detmodf} (a), where the modes enhanced by the detector anisotropies have been successfully suppressed.
	
	From Fig.~\ref{fig:mc2_detmodf} (b), the ratios for $\ell \leq 6$ stand within $1\sigma$ deviation. On the other hand, the error on the estimation of the higher modes is significantly wide, though still comparable to $MC_1$ itself. It has been mentioned that in this region the influence of $v_n$ wanes, making way for the detector anisotropies. So it is not surprising that this region has wide error bars. 
	
	Detector limited acceptance, event multiplicities and detector efficiency all have counter strategies. Then, with the method for angular power spectrum estimation established, it is time to move on to data.   	
	
	
	\section{Applying to data}
	
	The correction methods described in the previous section successfully returned the expected average power spectra values in the low $\ell$ regime, i.e. $\ell \leq 6$. Given this result, we may confidently apply the latest discussed method to data itself. The first consideration is vertex selection, a feature not dealt with in Ref.~\cite{PhysRevC.99.054910}. Then, the corrected angular power spectrum $\langle S^{m\neq0}_{\ell} \rangle$ for heavy-ion data is displayed for all centralities. The translation from power spectrum to flow is yet calculated again, since the resolution has changed. These values are also compared to $v_n$ computations with an $\eta$ gap. Lastly, the transverse momentum phase space is separated and each of their corresponding spectra calculated.
	
	\subsection{Vertex selection}
	
	When heavy ions collide in the ALICE detector, the resultant particles firstly reach the Inner Tracking System (ITS) of the experiment, a cylindrical detector whose main tasks consist in primary and secondary vertices reconstruction, aside from tracking and identification of particles~\cite{Collaboration_2008}. This subsection essentially focuses on the location of each event's primary vertex, or interaction point, and how the resulting power spectrum changes when selecting the position where the collision occurred and then following the steps depicted on previous sections.
	
	\begin{figure}[!ht]
		\includegraphics[width=0.4\textwidth]{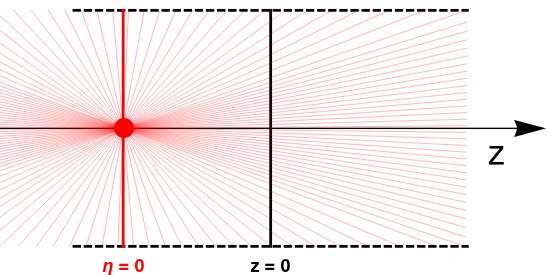}
		\caption{\label{fig:zvtx_img} Scheme of a heavy ion collision whose primary vertex is located at $z=-7.5\unit{~cm}$}
	\end{figure}	
	
\begin{figure}[!ht]
		\includegraphics[width=0.48\textwidth]{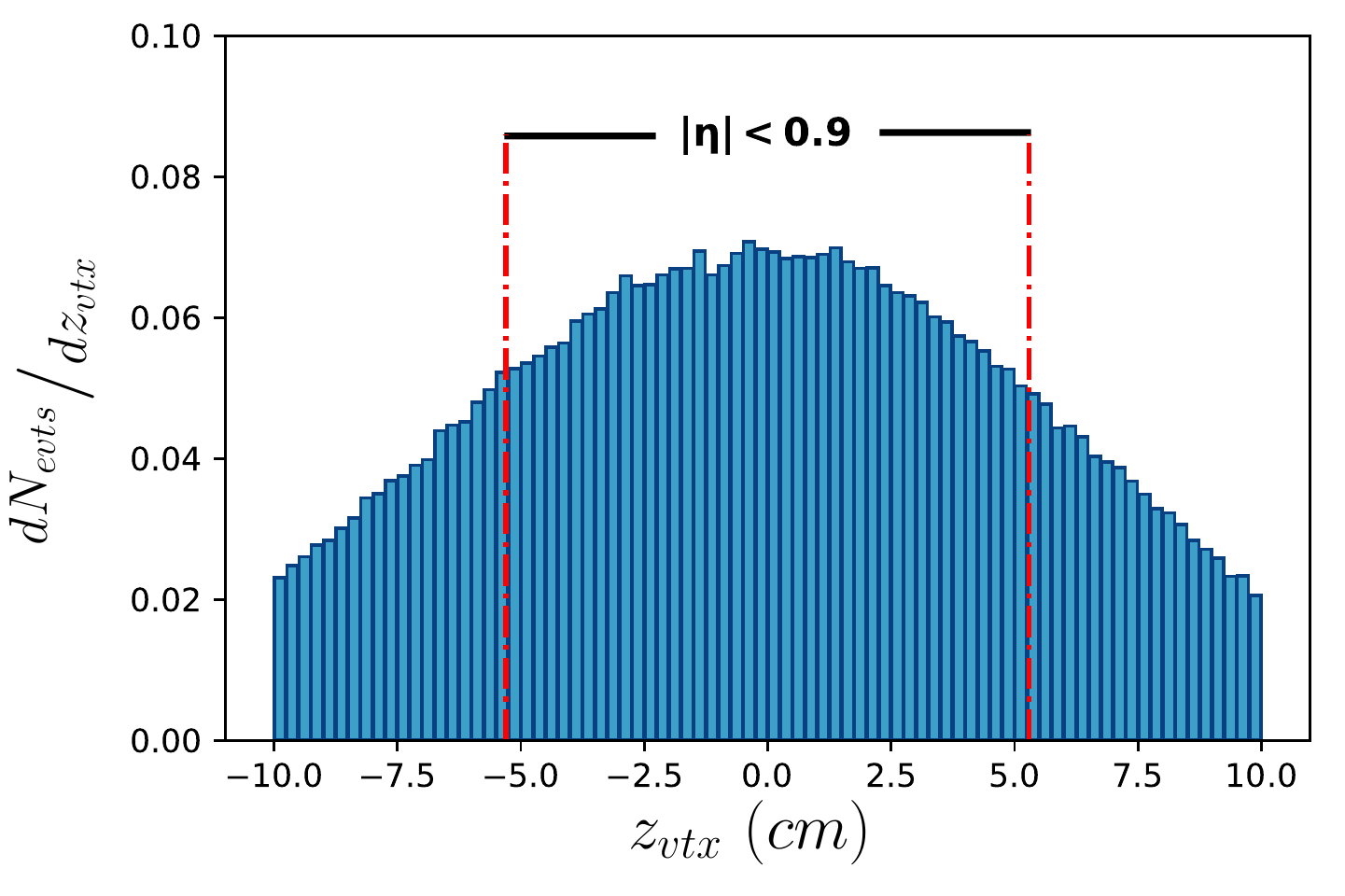}
		\caption{\label{fig:zvtx_dist} Distribution of $N_{evts}$ according to their $z_{vtx}$. The red lines indicate the $z$ position where acceptance is uniform for $|\eta| < 0.9$.}
	\end{figure}

	The interaction point of an event may be located anywhere along the beam axis $z$ spamming a couple of centimeters along the interaction region. For the data set at hand, each collision happened in the interval $-10\unit{~cm} < z < 10\unit{~cm}$, where $z = 0\unit{~cm}$ is in the center of the ITS detector, represented by a thick black line in Fig.~\ref{fig:zvtx_img}. The latter shows an schematic representation of the ITS' cross section. At mid-rapidity, i.e. $\eta = 0$, the collision happens: the lines coming out of it crudely represent the resultant particles, whose directions of emission spam over the whole pseudorapidity (or $\theta$) and azimuthal ranges. However, with the primary vertex located at, for instance, $z_{vtx}=-7.5\unit{~cm}$ and the detector's coverage limited to the dashed lines in Fig.~\ref{fig:zvtx_img}, particles with $0 > \eta > -0.9$ may end up not being tracked. Lastly, Fig.~\ref{fig:zvtx_dist} illustrates how events ($N_{evts}$) from the given run are distributed along the vertex: note how their bulk is located near the center, i.e. $z = 0\unit{~cm}$.
	
	Both in the previous work~\cite{PhysRevC.99.054910} and up until this point in the current study, we have performed the full analysis, from mapping particles on Mollweide projections to correcting $\langle C^{m\neq0}_{\ell} \rangle$ by multiplicity, on a batch of events whose interaction points were located anywhere along $-10\unit{~cm} < z < 10\unit{~cm}$. The objective of this section is then to first select in which interval with $dz = 2\unit{~cm}$ width the primary vertex is located. Secondly, the centrality division from 0-5\% to 30-40\% is determined and finally, the following steps are performed for each vertex interval: mapping particles, normalizing the event maps by $F^{all}(\mathbf{n_p})$ and calculating their power spectra.  
	
	\begin{figure}[!ht]
		\includegraphics[width=0.48\textwidth]{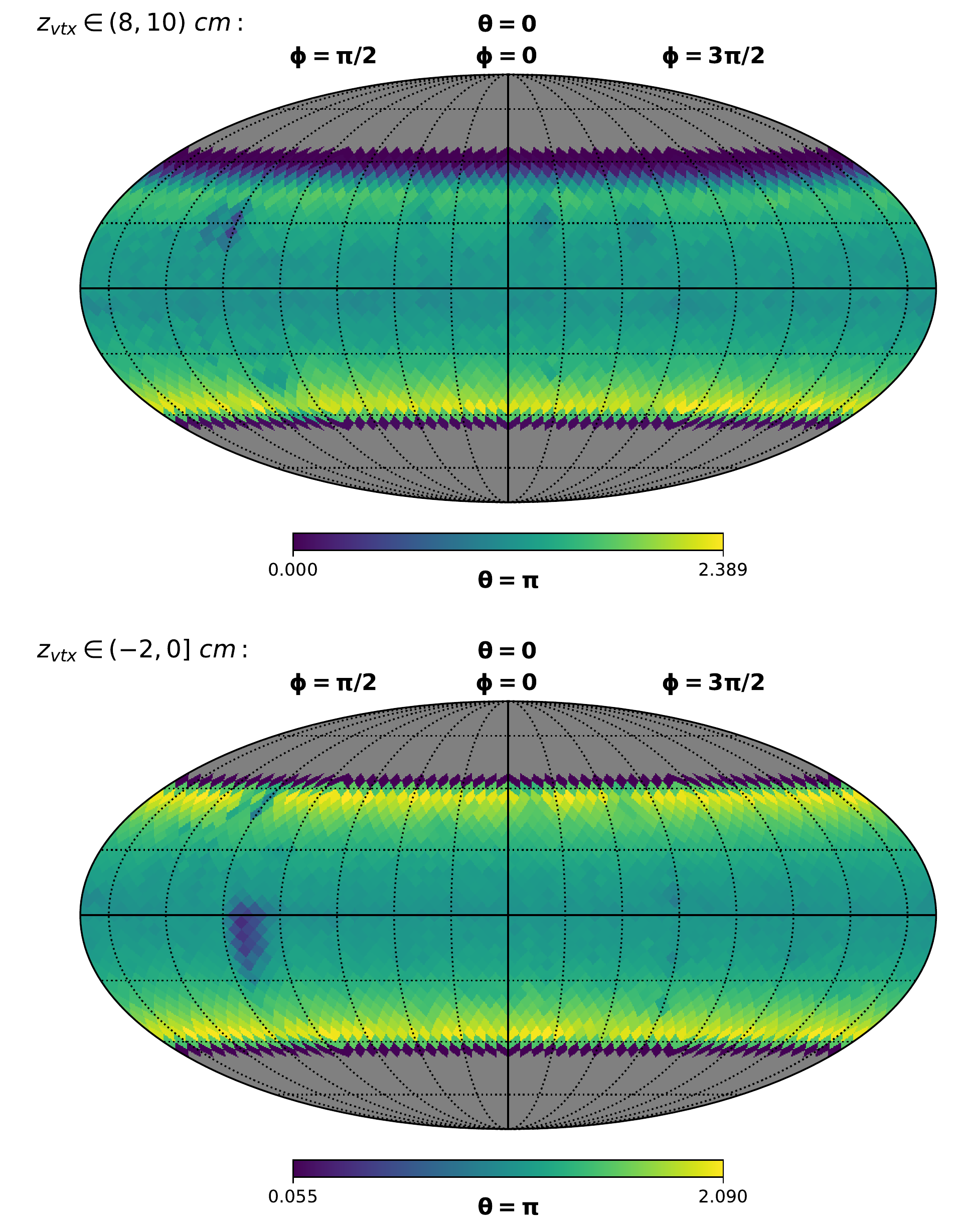}
		\caption{\label{fig:maps_zvtx02} $F^{all}(\mathbf{n_p})$ for events with primary vertex in $(8, 10)$ cm (top) and $(-2, 0]$ cm (bottom). Maps are from the 10-20\% centrality.}
	\end{figure}  	
	
	Events whose interaction points happen close to the detector's edge, end up with unaccounted particles whose pseudorapidity lies within $|\eta| < 0.9$. Those escape the detector at only one of its sides, as illustrated in Fig.~\ref{fig:zvtx_img}, which leads to an asymmetric $\theta$ distribution, shown on the top of Fig.~\ref{fig:maps_zvtx02}. This map is the average over all events with primary vertex located within $(8, 10)\unit{~cm}$: note its asymmetry as pixels close to the top edge are not colored yellow, contrasting with the ones on the bottom. 
	
	On the bottom of Fig.~\ref{fig:maps_zvtx02} is the average over all event maps whose interaction point lies within $(-2, 0]\unit{~cm}$ of the detector center. It is one of the regions where majority of events happen. In contrast to the map on top, both edges are colored yellow, as they possess a similar number of particles per pixel. At this point, it is important to remark that even though these events are from the same run, both $F^{all}(\mathbf{n_p})$ maps seem to have different detector-caused anisotropies. While for the one on top the inefficiency is located above $\theta = \pi/2$ ($\eta = 0$), for the other it is below. This suggests that the detector part responsible for the non-uniformity is located between the center and left edge of the interaction region. Unsurprisingly, the detector anisotropy seen in the mentioned maps also differ from the one seen on Fig.~\ref{fig:supmap1015}: as this map is a result of all events with primary vertices ranging from $-10\unit{~cm}$ to $10\unit{~cm}$ and the non-uniformity ends up smeared over $|\eta| < 0.9$.
	
\begin{figure}[!ht]
		\includegraphics[width=0.48\textwidth]{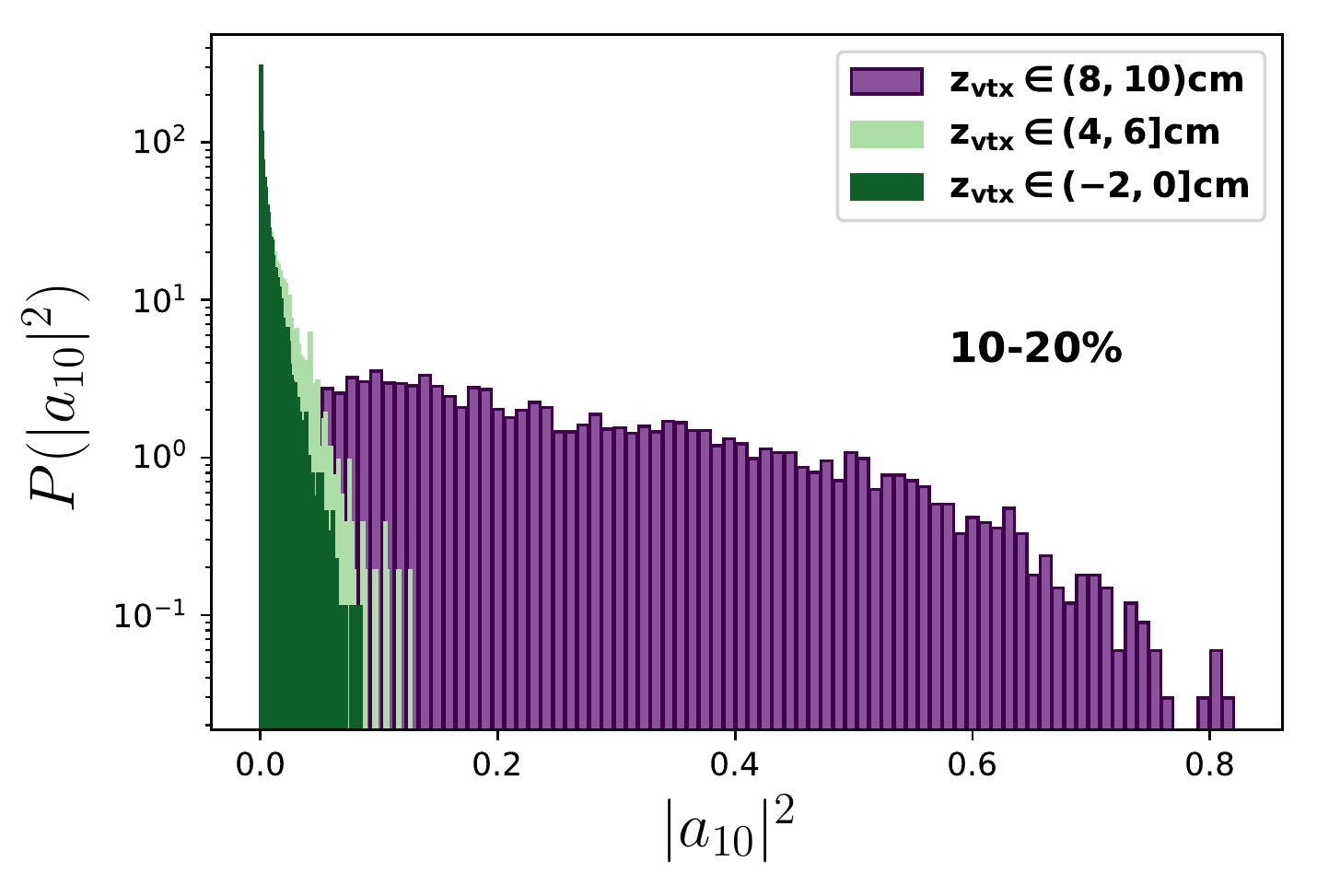}
		\caption{\label{fig:al0_sqr} Distribution of coefficients $|a_{1 0}|^2$ for events with different vertex intervals.}
	\end{figure}

	
	Due to the asymmetric geometry of the map from $(8, 10)\unit{~cm}$, $a_{\ell 0}$ modes with $\ell$ odd are expected to be non-trivial. In order to verify that, distributions of $|a_{10}|^2$ from events of three vertex intervals were created. Their results are compared in Fig.~\ref{fig:al0_sqr}, where it is clear to see that for $(8, 10)\unit{~cm}$, its $|a_{10}|^2$ values spam over three orders of magnitude. Meanwhile, for $(-2, 0]\unit{~cm}$ their average lies in $\mathcal{O}(10^{-3})$, the expected value for fully isotropic distributions with these multiplicity values.
	
	In the previous study~\cite{PhysRevC.99.054910} the full averaged power spectra $\langle C_{\ell} \rangle$ with $\ell$ odd for different centralities were found to follow a power law behavior $C(\ell) = A\cdot \ell^{-\beta} + \mathcal{C}$. Given how modes like $a_{1 0}$ are highly altered due to the asymmetries, this power law behavior of $\langle C_{\ell} \rangle$ for odd $\ell$ is simply a combination of asymmetric maps in $\theta$ and their $\phi$ anisotropy. 
	
	\subsection{Data spectrum}
	
	Finally we reach the point where we correct the spectra corresponding to each of the vertices, by subtracting $\langle N^{m\neq0}_{\ell} \rangle$ from $\langle C^{m\neq0}_{\ell} \rangle$. The $\langle C^{m\neq0}_{\ell} \rangle$ we are interested in correcting corresponds to that of normalized maps, $\overline{f}(\mathbf{n_p})$, whose anisotropies caused by detector efficiency should have been smoothed out. In this case, its averaged power spectrum should be subtracted by $\langle N^{m\neq0}_{\ell} \rangle_{bar}$, calculated from data maps azimuthally randomized and normalized by their averaged map. This step is repeated for all intervals at hand separately and, at the end, the average over the intervals is taken, a quantity denoted by $\langle S^{m\neq0}_{\ell} \rangle_z$. 
	
	\begin{figure}[!ht]
		\includegraphics[width=0.48\textwidth]{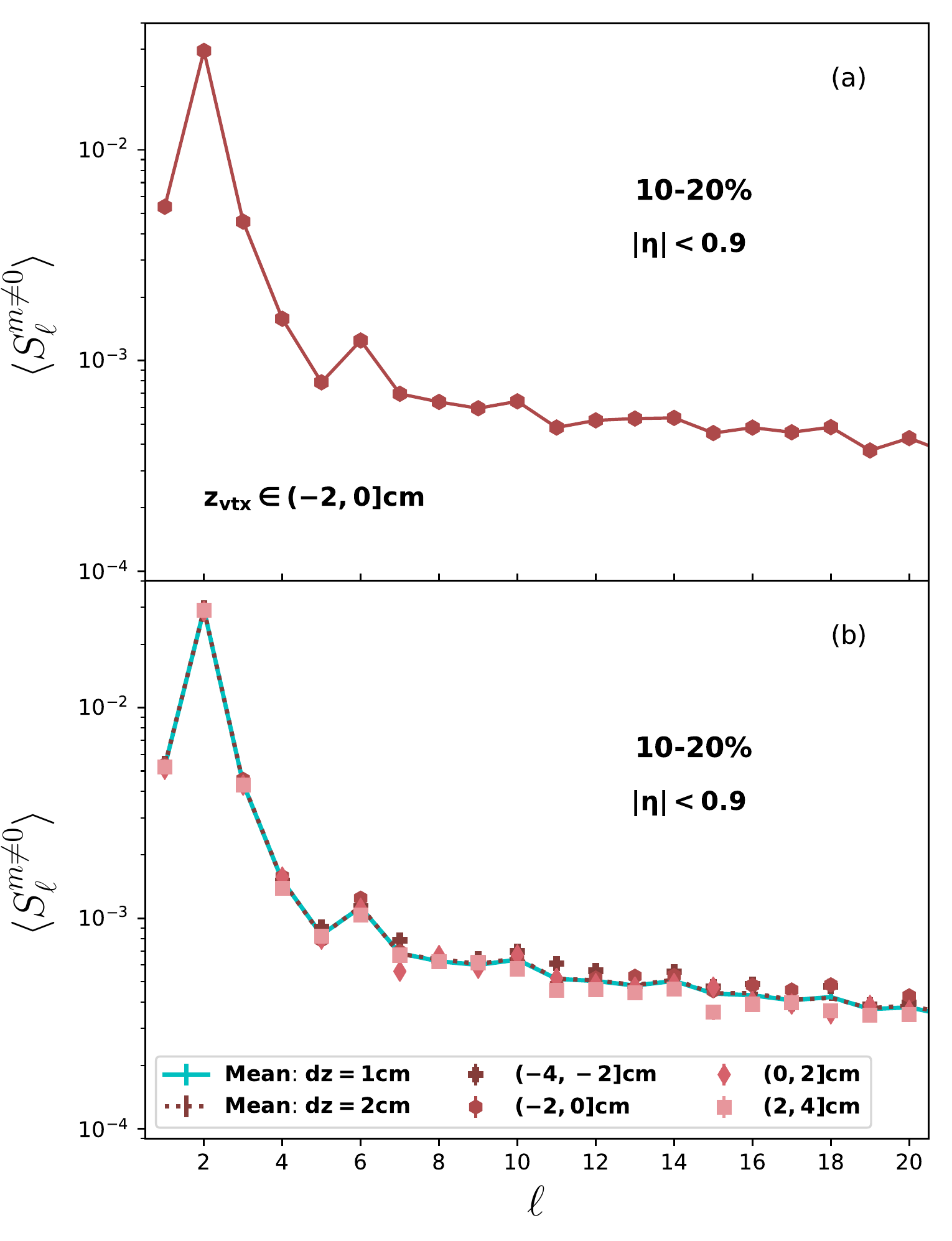}
		\caption{\label{fig:avgcls_zvtx_corr} Estimated $\langle S^{m\neq0}_{\ell} \rangle$ for the interval $(-2, 0]\unit{~cm}$ (a). Corrected average spectra for each vertex interval and $\langle S^{m\neq0}_{\ell} \rangle_z$ over all vertices for interval widths $dz = 1\unit{~cm}$ and $dz = 2\unit{~cm}$ (b).}
	\end{figure}
	
	In this study, particles emitted from events with $z_{vtx}$ within different intervals are considered to be drawn from the same distribution $f(\mathbf{\hat{n}})$, but affected by distinct detector efficiency functions $D_z(\mathbf{\hat{n}})$, where the subscript $z$ indicates that it is vertex dependent. For the current analysis, an interval width of $dz = 2\unit{~cm}$ was chosen to maximize the number of events per interval, while maintaining $D_z(\mathbf{\hat{n}})$ approximately the same for the sample. Additionally, an uniform acceptance within $|\eta| < 0.9$ is only possible for vertices $|z_{vtx}| < 5.3\unit{~cm}$~\cite{Collaboration_2008}. That explains why the intervals in Fig.~\ref{fig:avgcls_zvtx_corr} (b) range from $-4\unit{~cm}$ to $4\unit{~cm}$.  
	
	The corrected average power spectrum for the 10-20\% centrality and vertex interval $(-2, 0]\unit{~cm}$ is depicted in Fig.~\ref{fig:avgcls_zvtx_corr} (a). Notice that it possesses the same peak at $\ell = 6$ present in the MC simulated distributions. The higher peak at $\ell = 2$ is also unsurprisingly conserved, a possible sign of elliptic flow, as already mentioned. In addition, it has a damping tail with periodic `dips' on every fourth mode counting from $\ell = 8$ until $\ell = 20$. As seen in the MC cases, a spectrum fully dominated by flow anisotropies should drop significantly from $\ell = 7$. This means the aforementioned tail suggests the presence of anisotropies yet unaccounted for, as it does not possess, for example, the `dip' at $\ell = 8$ present on the MC simulations. Those modes are probably dominated by short-ranged non-flow effects, such as jet cones or resonance decays.
	
	The spectra (lines) at Fig.~\ref{fig:avgcls_zvtx_corr} (b) are weighted averages over all vertex intervals. They are presented alongside $\langle S^{m\neq0}_{\ell} \rangle$ for each of the intervals within $(-4, 4]\unit{~cm}$ with a $dz = 2\unit{~cm}$ width (markers). As a means of testing the efficiency of the power spectrum estimation method developed, $\langle S^{m\neq0}_{\ell} \rangle_z$ was also calculated for vertices within $(-5, 5]\unit{~cm}$ with $dz = 1\unit{~cm}$. It readily agrees with $\langle S^{m\neq0}_{\ell} \rangle_z$ for $dz = 2\unit{~cm}$, indicating the reliability of the method.
	
\begin{figure}[!ht]
		\includegraphics[width=0.48\textwidth]{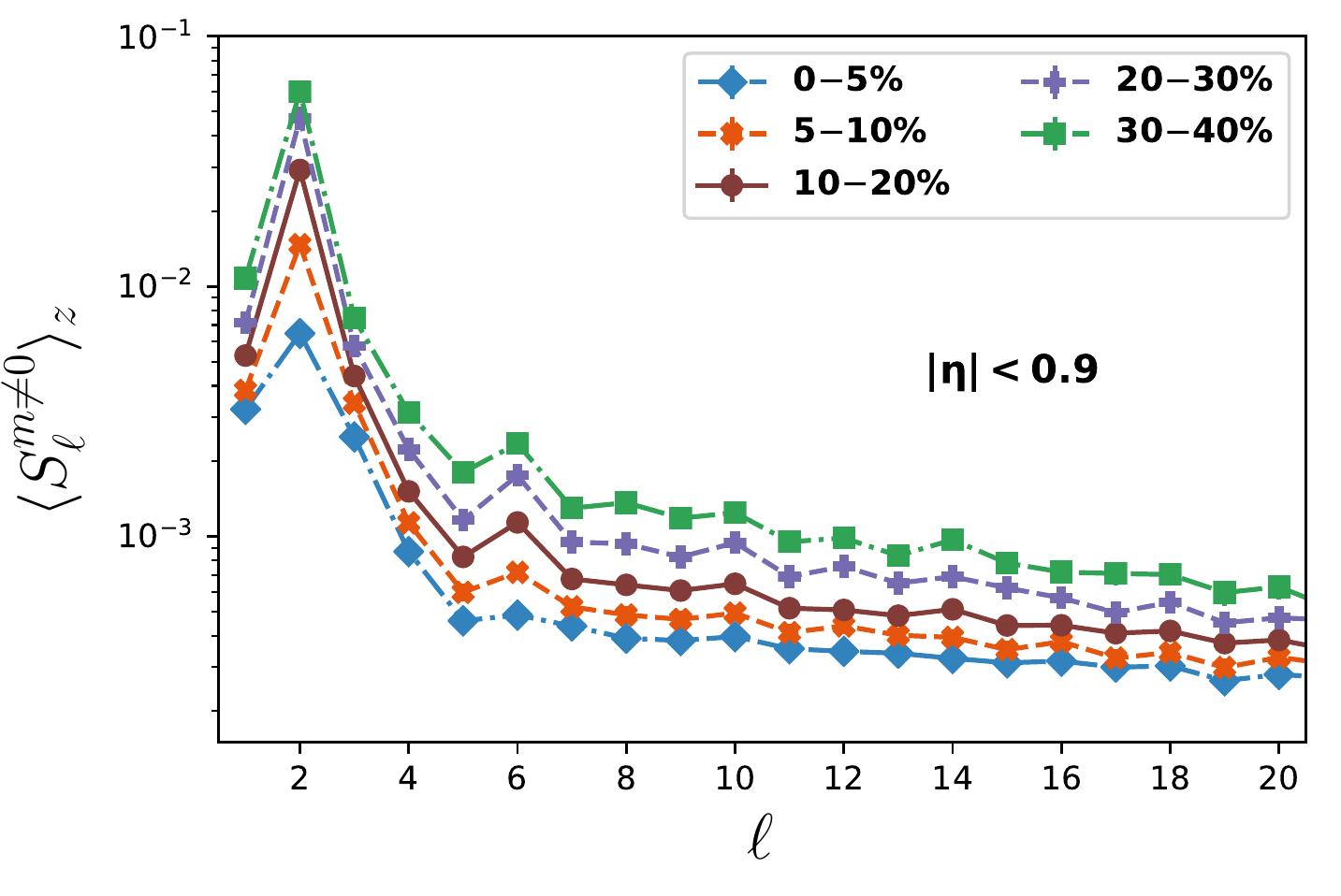}
		\caption{\label{fig:spec_all_centrs} Power spectra $\langle S^{m\neq0}_{\ell} \rangle_z$, $dz = 2\unit{~cm}$ for centrality intervals 0-5\%, 5-10\%, 10-20\%, 20-30\%, and 30-40\%.}
	\end{figure}

	Following up on the vertex averaged spectrum of 10-20\%, $\langle S^{m\neq0}_{\ell} \rangle_z$ for $dz = 2\unit{~cm}$ is calculated for the remaining centralities 0-5\%, 5-10\%, 20-30\%, and 30-40\%.  
	The current centrality division differs from the one in Ref.~\cite{PhysRevC.99.054910}, due to maximization of $N_{evts}$ after the separation in vertex intervals and $p_T$ phase space. The resultant $\langle S^{m\neq0}_{\ell} \rangle_z$ are shown in Fig.~\ref{fig:spec_all_centrs}. At first glance, the usual features can be spotted: a peak at $\ell = 2$ followed by decreasing values until a peak at $\ell = 6$. For higher modes, a damping tail dominates the spectra. There is also a clear hierarchy between centralities.  
	
	All the spectra in Fig.~\ref{fig:spec_all_centrs} have the large-scale characteristics of a scenario with anisotropic flow, since for $\ell \leq 6$ they resemble the MC spectra in Fig.~\ref{fig:mc_corr_avgcls}. Also, how their values increase with more peripheral centralities is akin to the behavior of azimuthal flow coefficients~\cite{alice_flow}. 
	
	The angular power spectrum describes correlations between $(\theta, \phi)$ pairs, so any phenomenon, from short to long-ranged, pertaining to two-particle correlations should contribute to $\langle S^{m\neq0}_{\ell} \rangle_z$. That includes jets, hadron decays, quantum fluctuations, among others. Since these are short ranged, their presence could be the main cause of the damping tail of $\langle S^{m\neq0}_{\ell} \rangle_z$.   
	
	\subsection{From spectrum to azimuthal flow}
	
	In light of the current approach to calculate the angular power spectrum of heavy-ions, it has been deemed important to repeat the azimuthal flow extraction from Ref.~\cite{PhysRevC.99.054910}. Not only has the resolution changed, but the events at hand were selected with uniform acceptance within $|\eta| < 0.9$. In other words, effects from maps asymmetric around $\theta = \pi/2$ were eliminated. Furthermore, $D_z(\mathbf{n_p})$ was considered for each vertex interval, thus assigning the corrected pixel weights from detector efficiency. We begin by finally presenting how azimuthal flow calculation works for MC simulations in comparison to the Q-cumulants method~\cite{qcumulants,qcumulants2} and then move on to the data spectra, where $v_n$ will be also computed with an $\eta$ gap. 
	
	From the flow ansatz, the azimuthal distribution of the emitted hadrons can be expanded in a Fourier series~\cite{flow1,flow2}:
	
	\begin{equation}
	\frac{dN}{d\phi} \propto \frac{1}{2\pi} \left[ 1 + 2\sum_{n=1}^{\infty} v_n \cos{n(\phi-\Psi_n)} \right],
	\label{eq:flow_ansz}
	\end{equation}  
	
	\noindent where  $v_n$ are the azimuthal flow coefficients and $\Psi_n$ are the symmetry planes associated with them. Consider now heavy ion events, from either simulations or real data, drawn from a factorizable distribution of the type $f(\mathbf{\hat{n}}) = g(\theta) h(\phi)$ as presented on the previous section. Additionally, take $h(\phi) = dN/d\phi$ as the flow ansatz in Eq.~\ref{eq:flow_ansz}. From the expansion of $f(\mathbf{\hat{n}})$ in spherical harmonics~\cite{PhysRevC.99.054910}:
	
	\begin{align}
	a_{\ell 0} &= b_{\ell 0} &\text{ for } m = 0\nonumber,\\
	a_{\ell m} &= b_{\ell m}\cdot v_{|m|} e^{-im\Psi_{|m|}} &\text{ for } m \neq 0,
	\label{eq:fourier_alms}
	\end{align}
	
	\noindent with 
	
	\begin{equation}
	b_{\ell m} = N_{\ell m} \int_{\theta_i}^{\theta_f} g(\theta) P_{\ell m}(\cos{\theta}) \sin{\theta} d\theta,
	\label{eq:blm} 
	\end{equation}
	
	\noindent where $(\theta_i, \theta_f)$ correspond to $\eta = 0.9$ and $\eta = -0.9$, respectively. Additionally, $N_{\ell m}$ is the square root coefficient of the spherical harmonics and $P_{\ell m}$ are the associated Legendre polynomials. 
	
	The azimuthal flow coefficients $v_n$ are extracted by combining Eq.(\ref{eq:fourier_alms}) with the expression in Eq.(\ref{eq:powspec_mdz}):
	
	\begin{align}
	|v_n|^2 &= \frac{2n + 1}{2} \cdot \frac{C^{m\neq0}_n}{|b_{nn}|^2} \cdot \frac{|b_{00}|^2}{C_0} \text{~~~~or}
	\label{eq:v1v2_expr} \\
	|v_{n}|^2 &= \frac{1}{|b_{nn}|^2}\bigg[ \frac{2n + 1}{2} \cdot C^{m\neq0}_n - \nonumber \\
	&\frac{2n - 3}{2} \cdot \frac{|b_{nn-2}|^2}{|b_{n-2n-2}|^2} \cdot C^{m\neq0}_{n-2} \bigg] \frac{|b_{00}|^2}{C_0},
	\label{eq:v3v4_expr}
	\end{align} 
	
	\noindent where $|b_{00}|^2/C_0$ is a normalization factor. The expressions in Eqs.(\ref{eq:v1v2_expr}, \ref{eq:v3v4_expr}) are valid for $n = 1,2$ and $n = 3,4$, respectively. It should also be remarked that the $v_n$ coefficients are calculated from the averaged corrected spectrum $\langle S^{m\neq0}_{\ell} \rangle$. 
	
	We begin by confirming whether or not they are valid for the simulated distribution drawn from $f_{MC}(\mathbf{\hat{n}})$, explicitly expressed by Eq.(\ref{eq:mc_dist}). The objective consists in recovering the input $v_n$ values from the power spectrum alone, here denoted as $v_n\{C_{\ell}\}$,  and compare them to the Q-cumulants' result for 2-particle correlations, denoted as $v_n\{2, QC\}$ or $v_n\{2\}$. The simulations have the same multiplicity as each of the centralities from 0-5\% to 30-40\% with $v_n$ input values that increase as collisions become more peripheral. Flow coefficients are also computed from $\overline{f}_{D_2}(\mathbf{n_p})$, as said maps are closer in similarity to the data ones.
	
	\begin{figure}[!ht]
		\includegraphics[width=0.42\textwidth]{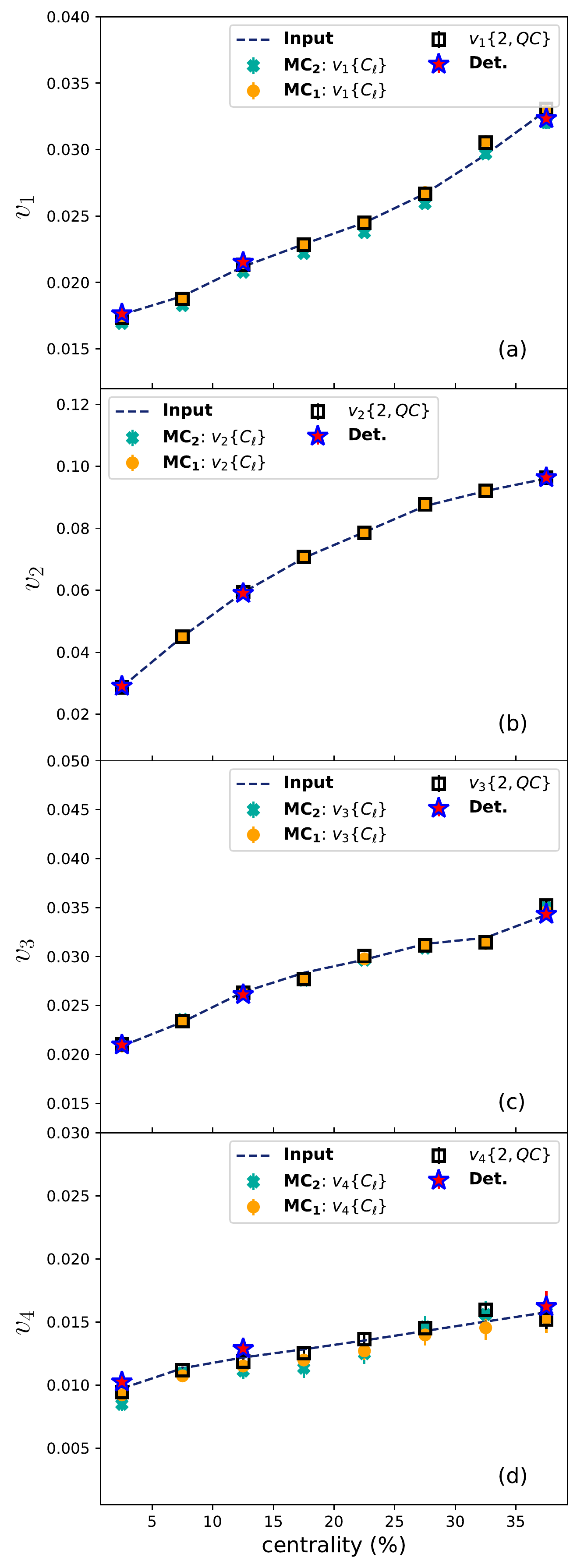}
		\caption{\label{fig:mc_vns} Comparison between the power spectrum extraction $v_n\{C_{\ell}\}$ and Q-cumulants $v_n\{2, QC\}$ to the input values of $v_n$, for $n=1,...,4$.}
	\end{figure}
	
	It is clear to see in Fig.~\ref{fig:mc_vns} that getting azimuthal coefficients from the power spectrum through Eqs~(\ref{eq:v1v2_expr}, \ref{eq:v3v4_expr}) works perfectly for the MC simulations. The resulting $v_n$ values are within error of the input ones and of the Q-cumulants' method. Even the coefficients from $\overline{f}_{D_2}(\mathbf{n_p})$ (stars) for centralities 0-5\%, 10-15\% and 35-40\% describe the input values. Given that these MC distributions are purely dominated by flow, the result is not surprising.
	
	
	As the extraction of flow coefficients through $\langle S^{m\neq0}_{\ell} \rangle$ works well for the MC-simulated emitted particles, the same steps were applied to the normalized maps of public data. The $v_n$ coefficients were calculated using Eqs.(\ref{eq:fourier_alms}, \ref{eq:blm}) for $b_{\ell m}$ from $g(\theta)$ as a constant, due to the normalization of data maps by their average $\theta$ distribution. This power spectrum estimation of azimuthal flow was compared again to the Q-cumulants method for two-particle correlations, now denoted $v_n\{2\}$. In addition to the latter, a pseudorapidity gap of $\Delta\eta > 1$ was imposed in order to suppress non-flow effects, this approach is denoted as $v_n\{2, \Delta\eta > 1\}$.  
	
\begin{figure}[!ht]
		\includegraphics[width=0.45\textwidth]{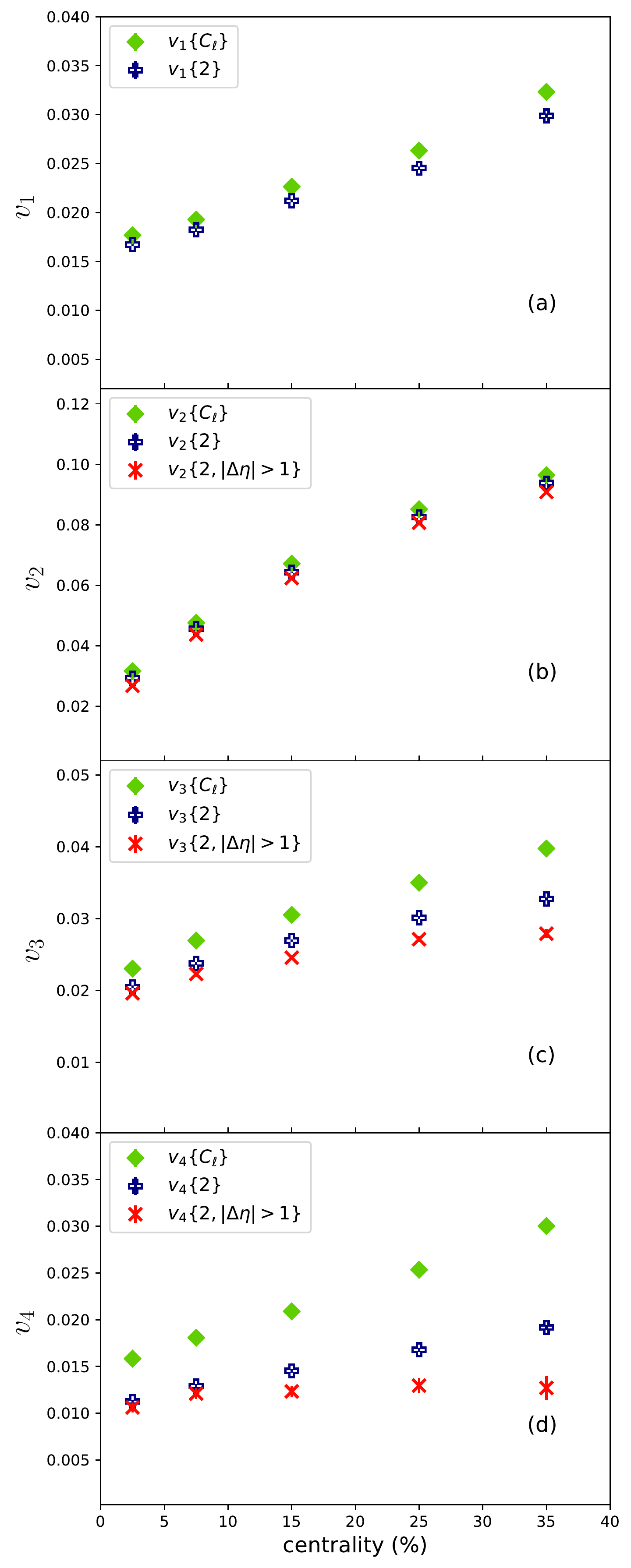}
		\caption{\label{fig:vns_comp} Comparison between power spectrum and Q-cumulants method for extraction of $v_n$ for data.}
	\end{figure}

	A trend can be observed in Fig.~\ref{fig:vns_comp}: flow coefficient values calculated using the power spectrum are typically higher than their Q-cumulants counterparts. This same effect had already been observed in Ref.~\cite{PhysRevC.99.054910} using $N_{side} = 8$ without accounting for the anisotropies from vertex selection. It is not surprising that $v_n\{C_{\ell}\}$ should stand above $v_n\{2, \Delta\eta > 1 \}$, since the latter does not consider short-ranged particle correlations in $\eta$. However, there is a difference between $v_n\{C_{\ell}\}$ and $v_n\{2\}$ despite the MC results suggesting that given a function $f(\mathbf{\hat{n}}) = g(\theta) h(\phi)$ they should be the same. In conclusion, the assumption that data follows a factorizable function even within $|\eta| < 0.9$ should be regarded carefully.
	
	Alternatively, consider the final event distribution as a superposition of maps: one with the large scale structures and another with the short-ranged ones. When computing the power spectrum, their final harmonic coefficients $|a_{\ell m}|^2$ would receive contributions not only from the single maps $|a^{A}_{\ell m}|^2$ and $|a^{B}_{\ell m}|^2$, but also from their cross-term $a^{A}_{\ell m}(a^B_{\ell m})^* + (a^{A}_{\ell m})^*a^B_{\ell m}$. Here $A$ and $B$ represent the large and small-scale structure maps.
	
	Yet another way of understanding the difference between $v_n\{C_{\ell}\}$ and $v_n\{2\}$ is to consider that the 3-D geometry of non-flow anisotropies is seen differently through spherical and azimuthal two-particle correlations. Lastly, the result of $v_n\{2, \Delta\eta > 1\}$ for $n = 1$ indicates that the values of $v_1$ are short-ranged in $\eta$, i.e., not an anisotropy related to a symmetry plane $\Psi_1$. For instance, single jets could contribute to the high dipole.
	
	\subsection{Transverse momentum}  
	
	In the previous sections, we showed that the average angular power spectrum from total multiplicity maps contains a non-trivial large-scale structure which is probably related to the extensively studied flow coefficients. It also has a damping tail, associated to short-ranged two-particle correlations. When comparing calculations of $v_n$ through $\langle S^{m\neq0}_{\ell} \rangle_z$ and the well known Q-cumulants method for the ALICE data, however, we found discrepancies between all flow harmonics, suggesting that $f(\mathbf{\hat{n}}) \neq g(\theta)h(\phi)$. 
	
	Correlations between produced particles in heavy ion collisions are also studied as a function of transverse momentum, $p_T$~\cite{alice_flow}. Additionally, the dependence of $v_n$ coefficients on $p_T$ also provides information on the hydrodynamical behavior of the QGP, aside from being sensitive to the medium's viscosity~\cite{azimuthal2}. In light of these, we study now how multiplicity maps and their spectra change with transverse momentum.
	
	The objective in this section is to make a simple, yet straightforward analysis of how the angular power spectrum changes with $p_T$. Therefore, we begin by separating the resulting particles of each event in two transverse momentum intervals: $p_T < 0.54\unit{~GeV}$, or \textit{lower $p_T$}, and $p_T > 0.54\unit{~GeV}$, or \textit{upper $p_T$}. The choice of such intervals results in each of their events having approximately the same multiplicity.
	
	After separating the particles of each event by their $p_T$ values, we perform the usual step of making their multiplicity maps by projecting their angular coordinates $(\theta_i, \phi_i)$ onto a sphere. Then a background spectrum $\langle N^{m\neq0}_{\ell} \rangle$ is estimated for both of them separately and subtracted from their spectra ensemble averages. Finally, the weighted average of the vertex spectra is taken, yielding $\langle S^{m\neq0}_{\ell} \rangle_{Uz}$ for $p_T > 0.54\unit{~GeV}$ and $\langle S^{m\neq0}_{\ell} \rangle_{Lz}$ for $p_T < 0.54\unit{~GeV}$.
	
	The maps under full momentum phase space studied so far can be seen as the superposition of a map whose particles have $p_T > 0.54\unit{~GeV}$ and another whose particles have $p_T < 0.54\unit{~GeV}$. The $a_{\ell m}$ of the full phase space map has the following relation to the harmonic coefficients of the aforementioned maps:
	
	\begin{equation}
	|a_{\ell m}|^2 = |a^U_{\ell m}|^2 + |a^L_{\ell m}|^2 + (a^U_{\ell m})^* \cdot a^L_{\ell m} + a^U_{\ell m}\cdot (a^L_{\ell m})^*,
	\label{eq:alms_pt}
	\end{equation}
	
	\noindent where $a^U_{\ell m}$ and $a^L_{\ell m}$ correspond, respectively, to the upper and lower bounds of the $p_T$ intervals.
	
\begin{figure}[!ht]
		\includegraphics[width=0.48\textwidth]{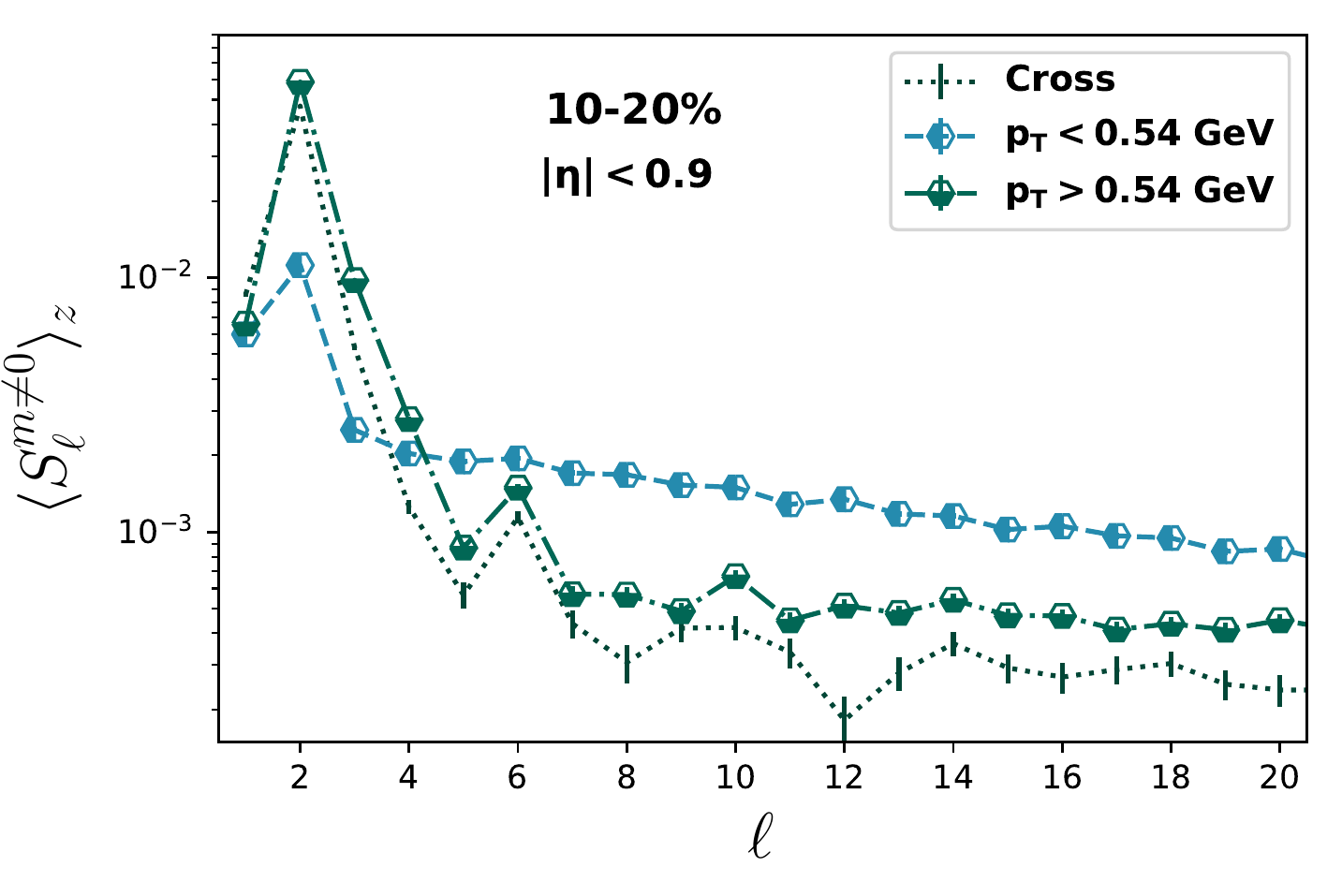}
		\caption{\label{fig:avpt_1020} Estimated $\langle S^{m\neq0}_l \rangle_z$ for lower, upper $p_T$ and their cross-term at 10-20\% centrality.}
	\end{figure}

	The terms in Eq.~(\ref{eq:alms_pt}) can be summed in $m$ for all $\ell$ and divided by $2\ell + 1$, showing that the spectra from Fig.~\ref{fig:spec_all_centrs} are equal to the sum of $\langle S^{m\neq0}_{\ell} \rangle_{Uz}$ and $\langle S^{m\neq0}_{\ell} \rangle_{Lz}$ plus a cross-term. Those three are depicted in Fig.~\ref{fig:avpt_1020} for the 10-20\% centrality. 
	
	Note from Fig.~\ref{fig:avpt_1020} how the spectra for $p_T > 0.54\unit{~GeV}$ and $p_T < 0.54\unit{~GeV}$ retained different characteristics after correction. Firstly, upper-$p_T$ has a more typical flow-dominated shape, with not only the peak in $\ell = 2$ and $\ell = 6$, but also $\ell = 10$, an effect highly influenced by $v_2$. Additionally, it has also a higher value at $\ell = 3,4$ than its counterpart. In turn, the power spectrum at lower-$p_T$ dominates the picture for $\ell \geq 5$ and it possesses a seemingly more slanted damping tail than the ones present in Figs.~\ref{fig:avgcls_zvtx_corr},~\ref{fig:spec_all_centrs}. 
	
	From Fig.~\ref{fig:avpt_1020}, it can be seen that particles with higher momentum encode most of the anisotropies arising from fluctuations in initial conditions, given how $\langle S^{m\neq0}_n \rangle_{Uz}$ for $\ell = 2, 3$ stand way above their lower momentum counterparts. This observation is in agreement with measurements of $v_n(p_T)$, which show that flow coefficients have higher values with increasing $p_T$~\cite{alice_flow}.
	
	The cross-term (dashed line) in Fig.~\ref{fig:avpt_1020} has similar shape compared to $\langle S^{m\neq0}_n \rangle_{Uz}$. Interestingly, the cross-term does not need correction for multiplicity, since the backgrounds of each $p_T$ spectrum are independent quantities, i.e., not correlated to each other.
	
\begin{figure}[!ht]
		\includegraphics[width=0.48\textwidth]{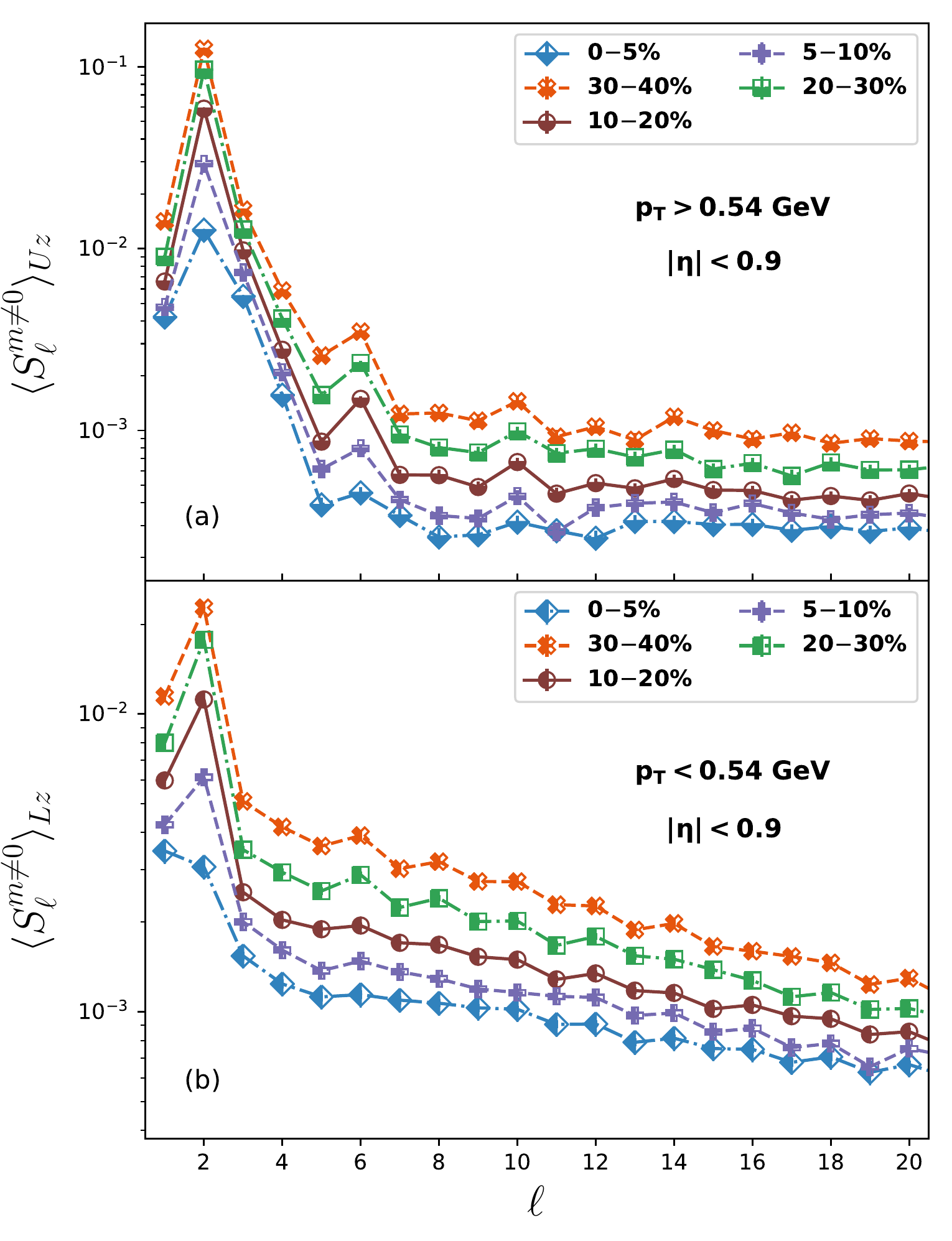}
		\caption{\label{fig:avpt_all_centrs} Angular power spectra of distributions with $p_T > 0.54\unit{~GeV}$ (a) and $p_T < 0.54\unit{~GeV}$ (b) as they change with centrality.}
	\end{figure}

	On the follow-up, the averaged spectra for upper- and lower-$p_T$ intervals are calculated for all centralities. The result is shown in Fig.~\ref{fig:avpt_all_centrs} for $p_T > 0.54\unit{~GeV}$ (a) and $p_T < 0.54\unit{~GeV}$ (b). Overall, the centrality hierarchy remains, with spectrum values increasing as collisions become more peripheral. Next, we tackle the characteristics of each spectrum separately.
	
	The features of $\langle S^{m\neq0}_{\ell} \rangle_{Uz}$ are quite sharp at $\ell \leq 6$, with values flattening out at higher $\ell$. The clear exception is $\ell = 10$, which suggests an increase in dominance of initial conditions anisotropies. The aforementioned plateau at higher $\ell$ indicates that the size of fluctuations does not change significantly with scale from $\ell = 11$.
	
	In the case of $\langle S^{m\neq0}_{\ell} \rangle_{Lz}$ it is now clear to see that the damping tails have become more slanted than in the full $p_T$ phase space spectra. Meanwhile, the peak at $\ell = 6$ has been smoothed out, though the one at $\ell = 2$ has remained for all but 0-5\%. The relatively high $\ell = 2$ suggests that these low momentum particles are globally arranged like the initial overlapping region. Said particles are associated with small-scale phenomena, which is hinted at by the surpassing $\ell = 1$ mode over $\ell = 2$ for 0-5\%. 
	
	Since the tails of $\langle S^{m\neq0}_{\ell} \rangle_{Uz}$ seem to decay with a power-law, a fit to the function $P_{ow}(\ell) = \mathcal{A}\cdot \ell^{\gamma}$ was performed to the modes with $3 \leq \ell \leq 20$ on the spectra of all centralities. Due to $\mathcal{A}$ being just a scaling factor, only the exponent $\gamma$ was plotted as a function of centrality.
	
	\begin{figure}[!ht]
		\includegraphics[width=0.44\textwidth]{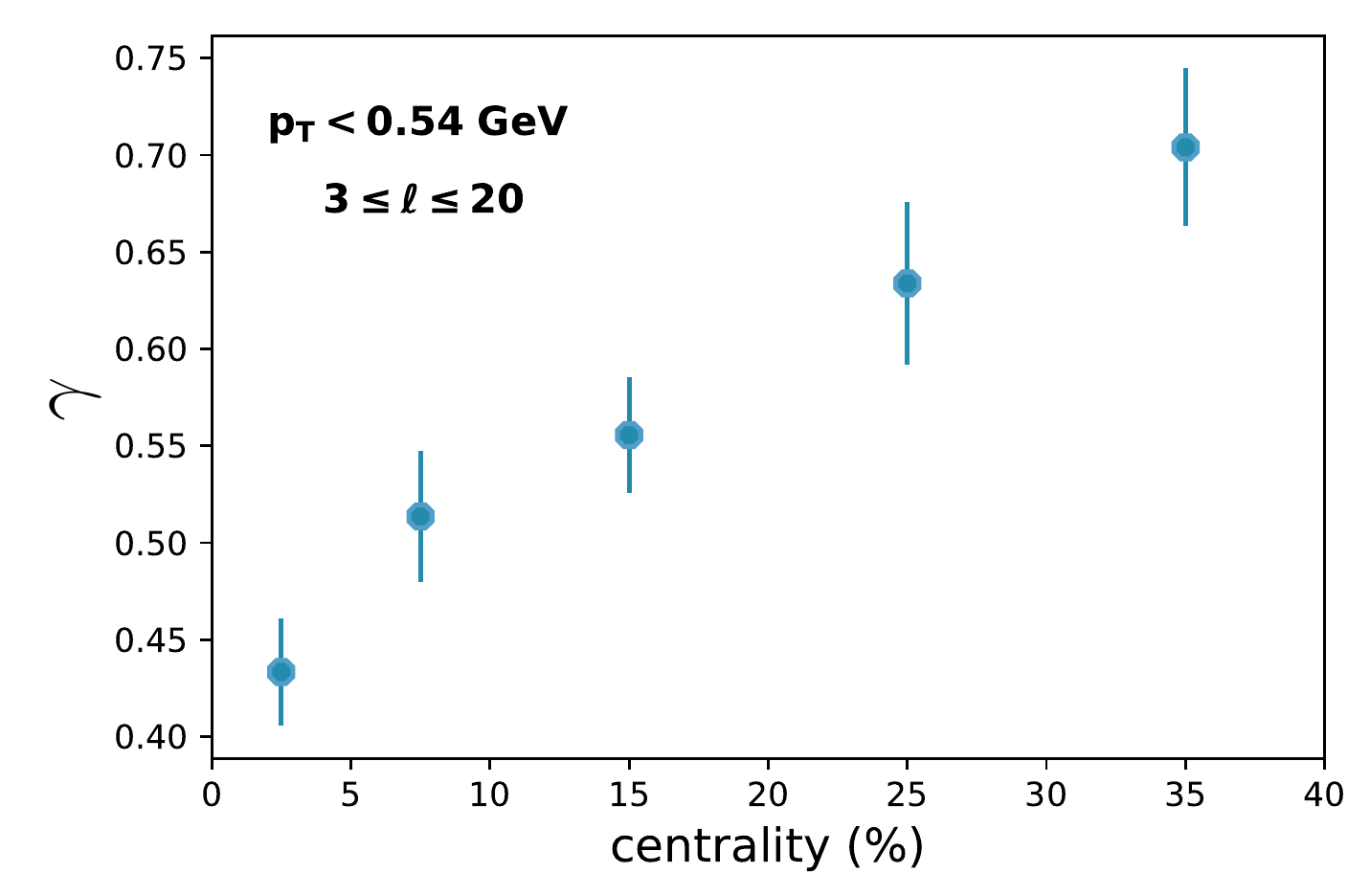}
		\caption{\label{fig:gamma} Power-law exponent $\gamma$ as it changes with centrality.}
	\end{figure}
	
	It can be observed from Fig.~\ref{fig:gamma} that $\gamma$ increases with centrality percentile. This means that the tails in Fig.~\ref{fig:avpt_all_centrs} (b) decrease faster for more peripheral collisions. One could consider the rarefaction of the medium: central events would have more particle clusters at smaller scales than peripheral ones, as their number of participants would allow for less particles being produced.
	
	Taking the geometry of momentum conservation to mean particles emitted on opposite sides of a sphere, it is possible to see that only $Y_{\ell m}$ with even $\ell$ have such symmetry. As a consequence, an spectrum influenced by momentum conservation would possess enhanced even modes. This effect can be slightly seen in the $\langle S^{m\neq0}_{\ell} \rangle_{Lz}$ of 20-30\% and 30-40\% centralities. The sparsity in the medium of events belonging to these centralities allow for particles to travel without interacting much after hadronization.
	
	The relatively high dipole moment ($\ell = 1$) persisted in all spectra seen in this work. This large-scale geometry is connected to $Y_{11}$ and $Y_{1 -1}$, which suggests a net asymmetry in $\phi$. Also, from the results of $v_n\{2, \Delta\eta > 1\}$, this anisotropy is short-ranged in $\theta$. A possible explanation could be a jet emitted on one direction, while its partner got swallowed by the medium.
	
	\subsection{Model comparison}
	
	The averaged power spectra over distinct vertex intervals $\langle S^{m\neq0}_{\ell} \rangle_z$ for both full-$p_T$ phase space and the intervals $p_T < 0.54\unit{~GeV}$ and $p_T > 0.54\unit{~GeV}$ have been displayed for heavy-ion data at $\sqrt{s_{NN}} = 2.76\unit{~TeV}$ measured with the ALICE detector. The results contain features associated with different sources of two-particle correlations, where global geometries arising from initial conditions dominate the low-$\ell$ region and phenomena unrelated to flow should be prominent in the high-$\ell$ region.
	
	Given characteristics such as the peaks in $\ell = 2$ and $\ell = 6$, as well as the damping tail from $\ell = 3$ in the spectra for $p_T < 0.54\unit{~GeV}$, a comparison to well-established models of heavy-ion collisions is a necessary step in this exploration of the angular power spectrum. Since spherical projections of data require knowing the final particle distribution both in the longitudinal and transverse directions, it is adamant to employ a 3+1D model. 
	
	A multi-phase transport model AMPT~\cite{ampt} explicitly handles non-equilibrium many-body dynamics. Overall, it consists of four main stages: initial conditions, parton dynamics, hadronization and hadronic interactions. Strings and minijets dominate the initial state, which is modeled with the heavy-ion jet interaction generator (HIJING)~\cite{hijing1,hijing2,hijing3,hijing4}. In the string melting~\cite{string_melt1,string_melt2,string_melt3} version of AMPT, excited strings are converted to partons according to their valence quarks. Then, the space-time evolution of the initial partons is treated with Zhang's parton cascade (ZPC)~\cite{zpc}. Subsequently, hadronization is described by a quark coalescence model which combines partons into hadrons. Finally, the latter's interactions are defined by a relativistic transport (ART)~\cite{art1,art2} model for hadrons. 
	
	The AMPT version employed in this study was \textit{v2.26t7b} with a string melting mechanism, released on May of 2018. The model was run for Pb-Pb collisions at center-of-mass energy per nucleon $\sqrt{s_{NN}} = 2.76\unit{~TeV}$. The choice of parameters followed Ref.~\cite{ampt2} and the resulting simulated particles reproduce mid-pseudorapidity and transverse momentum charged-particle distributions of ALICE heavy-ion data. Explicitly, the values for the screening mass and strong coupling constant are, respectively, $\mu = 2.265\unit{~fm^{-1}}$ and $\alpha_s = 0.33$, which correspond to a parton cross-section of $3\unit{~mb}$. In addition, the Lund string fragmentation parameters used are $a = 0.3$ and $b = 0.15\unit{~GeV^{-2}}$.  
	
	In this study, the AMPT power spectra are compared to ALICE data for a single centrality window, 10-20\%. The estimation of impact parameter values followed the expression $c \simeq \pi |\mathbf{b}|^2 / \sigma_{in}$~\cite{bformula}, where $c$ stands for centrality, $\mathbf{b}$ is the impact parameter and $\sigma_{in} \approx 784\unit{~fm^2}$ the nucleus-nucleus total inelastic cross-section calculated from the Glauber model. For the 10-20\% most central events, the minimum and maximum impact parameter values were thus $5.0\unit{~fm}$ and $7.1\unit{~fm}$, respectively.  
	
	Around $7000$ AMPT events were generated with random symmetry plane orientations for the 10-20\% centrality. Single event maps $f_{sim}(\mathbf{n_p})$ for each of these events were then created for particles with $|\eta| < 0.9$. Likewise experimental data, an all-event map $F^{all}_{sim}(\mathbf{n_p})$ was created from the overlap of all $\sim 7000$ $f_{sim}(\mathbf{n_p})$ maps. Since AMPT particle distributions are `measured by a perfect detector', $D_{sim}(\mathbf{n_p}) = 1$ and the all-event map is simply the average $\theta$-distribution, $F^{all}_{sim}(\mathbf{n_p}) = g_{sim}(\mathbf{n_p})$, in accordance to Eq.~\ref{eq:fall}. Following the steps of ALICE data spectrum we calculate $\bar{f}_{sim}(\mathbf{n_p}) = f_{sim}(\mathbf{n_p})/F^{all}_{sim}(\mathbf{n_p})$ in analogy to Eq.~\ref{eq:fbar}.
	
	The resulting maps $\bar{f}_{sim}(\mathbf{n_p})$ have assigned pixel weights according to the average $\theta$-distribution of AMPT events, which should agree with ALICE data given how both pseudorapidity distributions are consistent within $|\eta| < 0.9$. The angular power spectra of AMPT are then calculated from $\bar{f}_{sim}(\mathbf{n_p})$ and their average multiplicity background $\langle N_{\ell}^{m\neq0} \rangle_{sim}$ is estimated in the same manner as done for the ALICE data itself. Specifically, $10^6$ events following the AMPT $\theta$-distribution, $g_{sim}(\mathbf{n_p})$, and uniform in $\phi$ were generated. Their multiplicities reflected those of the AMPT events themselves. Each created event map was then divided by $g_{sim}(\mathbf{n_p})$ to assign pixel weights which match those of the AMPT maps. Their spectra were calculated and averaged over yielding $\langle N_{\ell}^{m\neq0} \rangle_{sim}$.
	
	\begin{figure}[!ht]
		\includegraphics[width=0.48\textwidth]{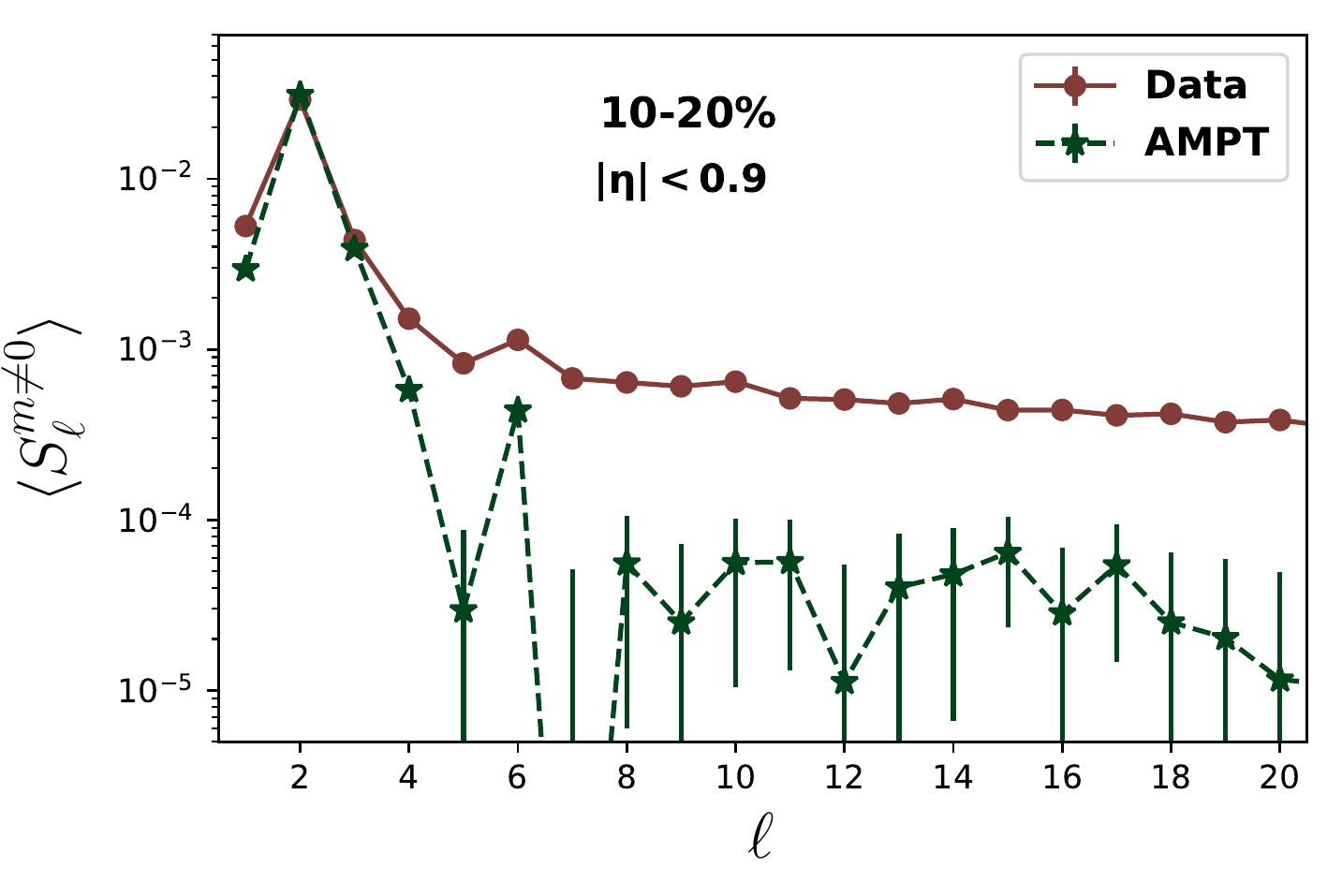}
		\caption{\label{fig:ampt_full} Comparison between averaged power spectra $\langle S^{m\neq0}_{\ell} \rangle$ from ALICE data and AMPT model.}
	\end{figure}
	
	The resulting averaged angular power spectrum of AMPT events $\langle S^{m\neq0}_{\ell} \rangle_{sim}$ is depicted in Fig.~\ref{fig:ampt_full}. Additionally, it is compared to the ALICE data spectrum $\langle S^{m\neq0}_{\ell} \rangle_z$ for the 10-20\% centrality. Note how the AMPT spectrum reasonably matches data for $\ell = 2,3$. More specifically, $\langle S^{m\neq0}_{\ell} \rangle_{sim}$ is within ~6.5\% and ~11.2\% of the data spectrum values for $\ell = 2$ and $\ell = 3$, respectively. 
	
	Given the aforementioned result, the next step consisted in estimating $v_n$ for AMPT using both Q-cumulants and $C_{\ell}$ - Eqs.~(\ref{eq:v1v2_expr},~\ref{eq:v3v4_expr}). Recall from Fig.~\ref{fig:vns_comp} that, in the ALICE data case, $v_n\{2\}$ and $v_n\{C_{\ell}\}$ diverge from each other specially for $\ell > 2$. As means of quantifying the differences between both $v_n$ calculations, we compute $|1 - v_n\{2\}/v_n\{C_{\ell}\}|$ for both the experimental data and AMPT. 
	
	The relative difference between Q-cumulants and $C_{\ell}$ on the ALICE data case more than doubles as $n$ increases, with $v_n\{C_{\ell}\}$ always yielding a higher value than $v_n\{2\}$; see Fig.~\ref{fig:vns_comp}. On the other hand, $v_n\{C_{\ell}\} < v_n\{2\}$ for $n = 3,4$ in the AMPT case. Additionally, the simulation's result $|1 - v_4\{2\}/v_4\{C_{\ell}\}| \approx 0.072 \pm 0.066$ was the highest among $n = 2,3,4$, with the others lying below $0.02$. In other words, both Q-cumulants with two-particle correlations and the power spectrum estimations of $v_n$ yield the same results in the AMPT case. A feature not present in the experimental data.
	
	All in all, the AMPT spectrum reproduces the shape of the ALICE data spectrum until $\ell = 6$. Beyond this multipole value, $\langle S^{m\neq0}_{\ell} \rangle_{sim}$ is, on average, a full order of magnitude below data and has a shape more indicative of an isotropic spectrum. An increase in number of events could aid in defining the shape better, though it is unlikely that it would match the data spectrum.
	
	\begin{figure}[!ht]
		\includegraphics[width=0.48\textwidth]{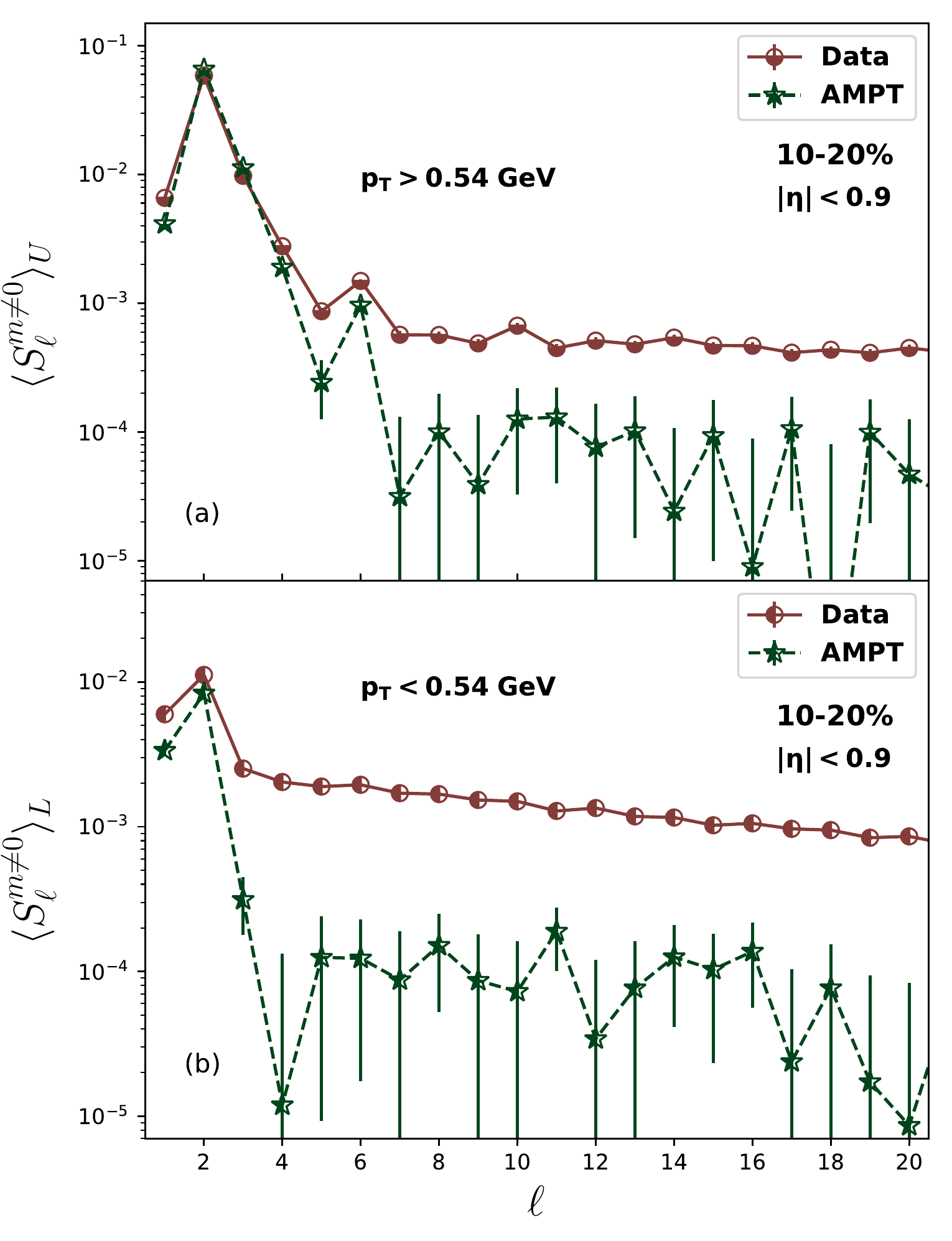}
		\caption{\label{fig:ampt_pt} Angular power spectra of ALICE data and AMPT model for $p_T > 0.54\unit{~GeV}$ (a) and $p_T < 0.54\unit{~GeV}$.}
	\end{figure}
	
	Likewise the full-$p_T$ phase space case above, AMPT angular power spectra are also calculated for the transverse momentum intervals $p_T > 0.54\unit{~GeV}$ and $p_T < 0.54\unit{~GeV}$. The events pertain to the same batch presented before, within the 10-20\% centrality class. We follow the same steps described before: create event maps, divide each by the total ensemble sum, calculate their spectra and take their average. In order to estimate the background, create $10^6$ events with $g_{sim}(\mathbf{n_p})$ as their $\theta$-distribution and uniform in $\phi$. Divide each event by $g_{sim}(\mathbf{n_p})$, calculate their spectra and average over them, finally yielding $\langle N_{\ell}^{m\neq0} \rangle_{sim}$ for each $p_T$ interval.
	
	The corrected spectra for $p_T > 0.54\unit{~GeV}$ and $p_T < 0.54\unit{~GeV}$ are shown in Fig.~\ref{fig:ampt_pt} (a) and (b), respectively. They are compared to the data spectra for 10-20\% already presented in Fig.~\ref{fig:avpt_all_centrs}. The AMPT power spectra for the upper and lower $p_T$ bounds shall be denoted $\langle S^{m\neq0}_{\ell} \rangle_U^{sim}$ and $\langle S^{m\neq0}_{\ell} \rangle_L^{sim}$, respectively. 
	
	In the case of $p_T > 0.54\unit{~GeV}$, a proximity of $\langle S^{m\neq0}_{\ell} \rangle_U^{sim}$ to the data spectrum for $\ell = 4,5,6$ is observed. Similarly to Fig.~\ref{fig:ampt_full}, AMPT describes the shape of the spectrum until $\ell = 6$. There is no peak at $\ell = 10$, a feature that becomes prominent when the transverse momentum phase space is sliced. Instead, for $\ell > 6$ $\langle S^{m\neq0}_{\ell} \rangle_U^{sim}$ stands closer to the MC cases above than experimental data. There is no coherent pattern, which suggests that at such scales, AMPT particle distributions are nearly isotropic.
	
	As for $p_T < 0.54\unit{~GeV}$, $\langle S^{m\neq0}_{\ell} \rangle_L^{sim}$ at $\ell = 2$ remains enhanced relative to the other multipoles, indicating the presence of an overall geometry connected to the initial overlapping region. On the other hand, higher $\ell$ possess the same behavior as previous spectra: no significant pattern and large error bars. This enforces the idea that AMPT within the current settings has no specific small scale structure. Meanwhile, the data spectra have a damping tail which becomes more slanted with peripheral collisions.
	
	Lastly, the AMPT spectra also have an azimuthal asymmetry short-ranged in $\eta$. This is further supported by the non-zero $v_1\{2\}$, the Q-cumulants estimation without $\eta$-gap.   
	
	
	\section{Discussion}
	
	This work explored two-particle correlations on a sphere, while it aimed at furthering our knowledge on anisotropies along the beam axis. The method presented in Ref.~\cite{PhysRevC.99.054910} was firstly detailed and applied to Monte Carlo simulated distributions in a pure flow scenario. Possible issues arising from limited detector acceptance and efficiency, as well as event multiplicity were properly tackled. Overall, the resulting spectra managed to describe the distributions up until an angular scale corresponding to $\ell = 10$.
	
	Once the method of power spectrum estimation had been established, it was applied to ALICE data. Considerations on the vertex position of each event had to be made, since for $|z_{vtx}| > 5.3\unit{~cm}$ acceptance for $|\eta| < 0.9$ is non-uniform and gives rise to artificial anisotropies; e.g. the power law behavior of uncorrected odd modes~\cite{PhysRevC.99.054910}. Given a vertex interval, events were classified in the following centralities: 0-5\%, 5-10\%, 10-20\%, 20-30\%, and 30-40\%. Their observed and background power spectra were calculated and their averages subtracted from each other, yielding $\langle S^{m\neq0}_{\ell} \rangle$. Lastly, the latter's weighted average over all vertices was taken, resulting in Fig.~\ref{fig:spec_all_centrs}.
	
	The most striking characteristic of $\langle S^{m\neq0}_{\ell} \rangle_z$ is the peak at $\ell = 2$, a clear signal of the initial almond shape imprinted in the final distribution. Additionally, the similarity to the MC spectra in a pure flow scenario indicates that primordial fluctuations are dominant in the region of large-scale structures, $\ell \leq 6$. The exception is $\ell = 1$, whose anisotropies are not related to a symmetry plane $\Psi_1$.
	
	Since $\langle S^{m\neq0}_{\ell} \rangle_z$ encompasses azimuthal anisotropies, $v_n$ were calculated using it through Eqs.~(\ref{eq:v1v2_expr},\ref{eq:v3v4_expr}) and compared to two-particle correlation calculations with and without an $\eta$ gap. Unsurprisingly, $v_n\{2, \Delta\eta > 1\}$ yielded lower values than the other two, due to suppression of non-flow effects. On the other hand, $v_n\{C_{\ell}\}$ remained higher than $v_n\{2\}$, suggesting that the assumption of data following a function $f(\mathbf{\hat{n}}) = g(\theta)h(\phi)$ might be insufficient, i.e., it could be true only on a first order approximation. Also, $v_n$ and $\Psi_n$ were taken to be approximately constant in $\theta$. Their variation with the polar angle could be a possible explanation for the difference between $v_n\{C_{\ell}\}$ and $v_n\{2\}$.
	
	The hierarchy $\langle S^{m\neq0}_2 \rangle_z > \langle S^{m\neq0}_3 \rangle_z > \langle S^{m\neq0}_4 \rangle_z$ is akin to that of $v_n$ coefficients with $n=2,3,4$. The increase of $\langle S^{m\neq0}_{\ell} \rangle_z$ values with centrality percentile is also similar to that of azimuthal flow coefficients. These are part of the body of evidence pointing towards geometries of initial conditions being imprinted on the angular power spectrum.
	
	It is with the exploration of the power spectra for different transverse momentum intervals that distinct geometries emerge. For instance, $\langle S^{m\neq0}_{\ell} \rangle_{Uz}$ possesses higher values for $\ell \leq 4$ in comparison to $\langle S^{m\neq0}_{\ell} \rangle_{Lz}$, while the latter dominates for $\ell > 5$. This can be explained due to $v_n(p_T)$ being higher for $p_T > 0.54\unit{~GeV}$ than for $p_T < 0.54\unit{~GeV}$. 
	
	The spectrum $\langle S^{m\neq0}_{\ell} \rangle_{Uz}$ has a peak in $\ell = 10$ in all centralities but 0-5\%, its values until $\ell = 6$ are also higher than the full spectrum. Additionally, its shape resembles more that of the presented MC spectra, strongly suggesting the influence of initial geometries in its features. Taking a look at the angular two-particle correlation functions $C(\Delta\phi, \Delta\eta)$~\cite{ATLAS:2012at}, small-range correlations $(\Delta\phi, \Delta\eta) \approx 0$ also influence $p_T > 0.54\unit{~GeV}$. The peak associated with these increases in value with centrality percentile. Also, it provides a reasonable explanation for the flattening out of the spectrum at $\ell > 10$, since the Fourier transform of a Dirac delta is a constant.
	
	For particles with $p_T < 0.54\unit{~GeV}$, $v_n(p_T)$ has relatively low values~\cite{alice_flow}, implying that for $\langle S^{m\neq0}_{\ell} \rangle_{Lz}$ the influence of primordial anisotropies dwindles significantly. Therefore, one could say that $\langle S^{m\neq0}_{\ell} \rangle_{Lz}$ is mainly a spectrum of non-flow. For instance, the enhanced even modes for $6 \leq \ell \leq 12$ in the centralities 20-30\% and 30-40\% suggest that $Y_{\ell m}$ with $\ell$ even have a considerable contribution to the spectrum. These are characterized by symmetries between points diametrically opposed, which could imply momentum conservation.
	
	The damping tail present on the CMB spectrum is caused mainly by photon diffusion, as these traveled from hot to cold areas of the universe, thus making it more uniform~\cite{silk_damp}. In other words, the scale of the fluctuations in the tail are comparable to the mean-free-path of photons. If the same interpretation is brought to the power spectra of heavy ions, then the observed suppression of higher $\ell$ modes indicates the length of mean-free-path $l_{mpf}$. What is more, $l_{mpf}$ would depend on the transverse momentum phase space, as the spectra for full-$p_T$, lower-$p_T$ and upper-$p_T$ differ from each other.   
	
	
	The comparison to AMPT showed that, under the current settings, the transport model managed to reproduce the shape of the data spectrum at $\ell \leq 6$ for both the full transverse momentum phase space and for $p_T > 0.54\unit{~GeV}$. The enhancement of the dipole ($\ell = 1$) and quadrupole ($\ell = 2$) moments is present on all three AMPT spectra, with the first indicating an overall asymmetry in $\phi$, though unrelated to a symmetry plane $\Psi_1$. As for the second, it consists of an imprint of the almond-shaped geometry generated by the overlapping nuclei.
	
	Despite the similarities described above, AMPT fails to emulate the data spectra for $\ell \geq 7$, in the cases of $\langle S^{m\neq0}_{\ell} \rangle_{sim}$ and $\langle S^{m\neq0}_{\ell} \rangle_U^{sim}$. For $p_T < 0.54\unit{~GeV}$, the discrepancy starts at $\ell = 3$, as AMPT completely lacks a damping tail. All in all, beyond these scales dominated by initial stage fluctuations, or flow, the AMPT spectra values have no distinct feature. One could say the AMPT particle distributions look isotropic when probed at smaller scales.    
	
	Overall, AMPT underestimates the size of fluctuations on particle distributions. Specially when it comes to short-ranged scales, where AMPT's particles seem to be nearly isotropic, i.e., they follow no specific pattern. In other words, the AMPT model results in smoother particle distributions, while experimental data is lumpier. 
	
	Considering the possible relation between the heavy-ion angular power spectrum and the system's mean-free-path, AMPT could have a larger $l_{mpf}$ than the experimental data. Specifically, one could think that the propagation of anisotropies to the final state is related to the mean-free-path. If the probed scale is bigger than $l_{mpf}$, than its patterns should be imprinted in the final distribution. However, if the opposite is true and $l_{mpf}$ is actually larger than the scale in consideration, then the latter's characteristics will be smoothed out. Hence the AMPT spectra being more suppressed relative to the ALICE data.   
	
	
	Given the comparison to the AMPT model and possible relations to the mean-free-path of the system, it would be interesting to submit results from hydrodynamic 3+1D simulations to the power spectrum analysis present in this study. One could also verify how different stages of the QGP creation and evolution affect the angular power spectrum.

	One of the main limitations of the developed method lies in the event multiplicities. With less particles, resolution should decrease, thus making the calculation of $a_{\ell m}$ less accurate. Analysis in the momentum range $p_T > 2\unit{~GeV}$, or of different particle species and small collision systems are highly limited. Precisely due to the latter, the task of separating the different causes of anisotropies could be challenging.
	
	This work adds to previous ones~\cite{powspec1,powspec2,powspec3} in power spectrum analysis of heavy-ion collisions and it differs primarily from them due to its thorough exploration of modes with $m \neq 0$. It also tackles the relation between angular power spectrum and transverse momentum differently than Ref.~{\cite{powspec2}}. In the latter, the pixels on the map correspond to $p_T$ values themselves, with the power spectrum measuring correlations between $\Delta p_T/p_T (\mathbf{\hat{n}})$ pairs. In the present work, the spectra at distinct $p_T$ intervals still measures correlations between $(\theta, \phi)$ pairs. Also, a depression in $\ell = 6$ was observed in the spectra of Ref.~{\cite{powspec2}}, while $\langle S^{m\neq0}_{\ell} \rangle_z$ displays the exact opposite, an influence of the initial almond-like geometry. Additionally, no acoustic peaks were present in the final spectra.
	
	In the current Big Bang paradigm, the quark-gluon plasma is the state of matter permeating the universe right after inflation. Curiously, the scale of QGP anisotropies is considerably larger ($\ell \leq 20$) than the CMB ones ($\ell > 100$), even thought the size of the universe at recombination is much larger than the size of the QGP droplet in LHC and RHIC. While the peaks of the heavy-ion spectrum tell of initial anisotropies, the CMB peaks are related to the curvature of the universe and matter densities.     
	
	\begin{acknowledgments}
		
		I would like to thank Poul Henrik Damgaard for the discussions and incentive on writing this paper. I am also grateful to Christian Bourjau for bringing to the table the possible issue of vertex selection. Finally, I thank Ante Bilandzic, J. J. Gaardh\o je, Pavel Naselsky, Hao Liu and You Zhou for the interesting discussions. This work was partly supported by the Danish National Research Foundation (DNRF) and the Conselho Nacional de Desenvolvimento Cient\'{i}fico e Tecnol\'{o}gico (CNPq).
		
	\end{acknowledgments}
	
	\nocite{*}
	
	\bibliography{paper2}
	
\end{document}